\documentstyle[psfig,draft]{mn2e}

\title[Radiation Transfer in PNe]
      {Radiation Transfer in the Cavity and Shell of Planetary Nebulae}

\author[M.\,D.\ Gray, M.\ Matsuura et al.]
       {M.\,D.\ Gray$^{1}$, M.\ Matsuura$^{2}$, A.\,A.\ Zijlstra$^{1}$\\ 
        $^{1}$ Jodrell Bank Centre for Astrophysics, Alan Turing
	       Building, University of Manchester, M13 9PL, UK\\
        $^{2}$ UCL-Institute of Origins, 
               Department of Physics and Astronomy, UCL,
               Gower Street,\\ London WC1E 6BT, UK}

\date{Accepted ... .
      Received ... ;
      in original form ...}

\pagerange{000--000}

\begin{document}

\maketitle

\begin{abstract}
We develop an approximate analytical solution for the transfer of
line-averaged radiation in the hydrogen recombination lines for the
ionized cavity and molecular shell of a spherically symmetric
planetary nebula. The scattering problem is treated as a perturbation,
using a mean intensity derived from a scattering-free solution.
The analytical function was fitted to H$\alpha$ and H$\beta$ data
from the planetary nebula NGC6537. The position of the maximum 
in the intensity profile produced consistent values for the radius
of the cavity as a fraction of the radius of the dusty nebula:
0.21 for H$\alpha$ and 0.20 for H$\beta$. Recovered optical depths were
broadly consistent with observed optical extinction in the nebula,
but the range of fit parameters in this case is evidence for a clumpy
distribution of dust.
   
\end{abstract}


\section{Introduction}
\label{txt_intro}

Shells with internal cavities are found in almost all PNe
\cite{Balick02}.
Such 
shells and cavities are related to the shaping of PNe by stellar winds.
Low and intermediate mass stars lose their mass 
at a very high rate, forming a circumstellar
envelope. Fast stellar winds from the central stars of  PNe overtake
the slower, denser, AGB ejecta, resulting in interaction of 
the fast and slow stellar wind components \cite{Kwok78}.
The fast wind sweeps the slow wind, and leaves a cavity inside.
Cavities are also found in supernova remnants, 
\cite{Dwek87,Lagage96}
but these structures are more complicated \cite{Bilikova07}. Shell-cavity
dimorphism is also found in LBV stars and in AGN.

 In one of the well studied bipolar planetary nebulae, NGC\,6537, 
the core consists of bright arcs tracing a shell surrounding
 an elongated cavity
\cite{Matsuura05}.
Arcs are bright in H$\alpha$ and other recombination lines, 
and extinction maps derived from H$\alpha$ and H$\beta$
suggests the presence of dust grains in the arcs.
 Matsuura et al.
\shortcite{Matsuura05} suggested that little dust exists in the cavity.
Deriving accurate dust density from H$\alpha$ and H$\beta$
line maps is not straightforward, as the light from the central star
is scattered within the arcs. In these arcs, both dust grains and
 gas are mixed together.
Therefore, we developed a radiative transfer code
to resolve cavity and shells for PNe, which includes scattered 
light in a shell.

A self-consistent radiative transfer code for this mixture has been developed
by Ercolano, Barlow \& Storey
\shortcite{Ercolano05}. They have used a Monte Carlo method.
Here, we use an analytical solution, with some approximations,
and concentrate on the configuration of 
cavities and shells around them. The aim of the paper is therefore
to obtain detailed physical insight through a simplified
analytical model, which can complement the more complex
information from
numerical solutions.

\section{The Model of NGC6537}

In this section, we describe the physical and radiative transfer
models that are used in our analysis of NGC~6573. Although certain
parameters are definitely specific to this object,
taken from Matsuura et al. \shortcite{Matsuura05},
much of the
model is generally applicable to any planetary nebula of similar
geometry, and in a similar state of evolution.

\subsection{Physical Model of the Nebula}
\label{txt_pmodneb}

\begin{figure}
\vspace{0.3cm}
\psfig{file=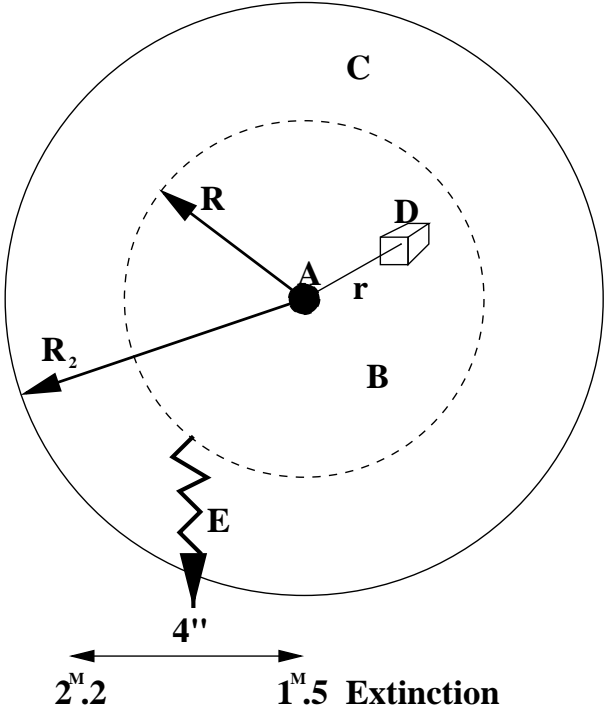,angle=0,width=8cm}
\caption{
A diagram of the planetary nebula model. The spherically symmetric
nebula is centered on the star, A, of negligible solid angle. This
is surrounded by the cavity, B, and molecular shell, C. An
infinitessimal
volume, D, of the cavity is marked at radius, \( r \). Optical
radiation, E, escapes from the shell after suffering some
degree of extinction. The particular values shown (in magnitudes) apply
to NGC6537.
}
\label{modelneb}
\end{figure}

We assume that the nebula NGC6537 is accurately spherically symmetric.
The layout of the various radial zones of the object is summarised
in Fig.~\ref{modelneb}. The central star is a white dwarf of
negligible solid angle, both from an observer's point of view, and
from the modeller's: no rays in the radiative transfer model are
considered to start or end on the stellar surface.

The central star emits sufficient vacuum ultraviolet radiation to
ionize a surrounding cavity. Ions in this cavity undergo frequent
recombination and photoionization cycles, and it is the main source
of H\(\alpha\) and H\(\beta\) line radiation. For NGC6537
specifically,
the observed flux ratio \( F_{H\beta}/F_{H\alpha} = 2.79 \), where
these fluxes are averages over the appropriate spectral-line bandwidth.
From this ratio, we adopt radially constant values of the
electron temperature, \( T_{e}=1.5\times 10^{4}\)\,K, and electron
number density, \(n_{e}=10^{4}\)\,cm\(^{-3}\).

Surrounding the H{\sc ii} cavity is a neutral shell, composed
mainly of molecular material. This shell contains dust, which we
assume, for the case of NGC6537, to have a mass fraction of
\( 0.01 \). Most of the shell material is, of course, in the
form of molecular hydrogen. The shell produces approximately
two magnitudes of visual extinction in NGC6537, derived
from the measured excess, \( E({\rm H}\alpha - {\rm H}\beta )\). The shell is
spatially resolved, and the extinction varies from 
\(1.5\)\,magnitudes at the centre to \(2.2\)\,magnitudes
 at \( 4\arcsec \) off
centre.

At present, we leave the form of the density profile as an
unknown function in both the cavity and shell zones of the
nebula.

\subsection{Radiative Transfer Model}

The volume element D in Fig.~\ref{modelneb} has volume \( \delta V \).
If we let this be unit volume, then the radiated power per unit
frequency and solid angle in the H\(\alpha\) line is
\begin{equation}
j_{\nu}(H\alpha )=\frac{h\nu_{32}}{4\pi} A_{32} n_{3}(r) \phi_{\nu , 32},
\label{emalpha}
\end{equation}
where \( \nu_{32} \) is the `laboratory' line-centre frequency of the
H\(\alpha\) line, and \( A_{32} \) is the Einstein coefficient for
spontaneous emission. The subscripts refer to the principal quantum numbers
of atomic hydrogen, so \( n_{3}(r) \) is the number density of H-atoms
in the upper state of the transition. With an isothermal
approximation,
the lineshape function, \( \phi_{\nu , 32} \) is not a function of
radius, but may be a function of direction if there are significant
radial motions in the cavity or shell. Given these definitions, a
similar expression may be written down for the emission coefficient
of H\(\beta\):
\begin{equation}
j_{\nu}(H\beta )=\frac{h\nu_{42}}{4\pi} A_{42} n_{4}(r) \phi_{\nu , 42}.
\label{embeta}
\end{equation}
We assume complete velocity redistribution, so that absorption
lineshape functions are identical to those specified above for
emission.

The radiative transfer equation is
\begin{eqnarray}
%\bmath{\hat{n}} \cdot \bmath{\nabla} I_{\nu} & = &
%-[\kappa^{c}(r) + \sigma^{c}(r) + \kappa_{\nu}(r)] I_{\nu} \nonumber \\
%& + & \kappa^{c}(r) B_{\nu}(T(r)) + j_{\nu}(r) \nonumber \\
%& + & \sigma^{c}(r) \oint \frac{d\Omega'}{4\pi}
%                \int_{0}^{\infty} d\nu' \Xi (\nu',\bmath{\hat{n}}';\nu,
%\bmath{\hat{n}}) I_{\nu'},
\bmath{\hat{n}} \cdot \bmath{\nabla} I_{\nu} & = &
-[\kappa^{c}(r) + \sigma^{c}(r) + \kappa_{\nu}(r)] I_{\nu} \nonumber \\
& + & \kappa^{c}(r) B_{\nu}(T(r)) + j_{\nu}(r)
+ \sigma^{c}(r) J_\nu (r),
\label{rt0}
\end{eqnarray}
where \( \bmath{\hat{n}} \) is a unit vector along the ray of specific
intensity \( I_{\nu} \). Opacity is provided by radius-dependent
absorption coefficients, \( \kappa_{\nu} \), \( \kappa^{c} \), for
the line and continuum, respectively, and the continuum scattering
coefficient, \( \sigma^{c} \). Kirchoff's Law allows the continuum
emission from dust to be written as \( \kappa^{c}(r) B_{\nu}(T(r)) \),
whilst the line emission coefficient is \( j_{\nu}(r)\), and will
be one of the specific forms in eq.(\ref{emalpha}) or
eq.(\ref{embeta})
for the appropriate transition. The final term on the right-hand
side of eq.(\ref{rt0}), equal to \( \sigma^{c}(r) J_{\nu} \), is the
scattering integral assuming isotropic, elastic scattering, and
$J_\nu$ is the angle-averaged intensity.

We expand the left-hand side of eq.(\ref{rt0}) in spherical polar
coordinates (for example Peraiah \shortcite{per02}). This
introduces \(\mu = \cos \theta \) where \( \theta \) is the
angle between the direction of ray propagation and the radial
direction. On the
right, we combine the continuum absorption and scattering into
an extinction coefficient, \( \chi^{c}(r) \), and
assume that the line absorption is negligible when
compared to that in the continuum. 
%Assuming that the
%scattering is elastic and isotropic, the complicated scattering
%integral reduces to \( \sigma^{c}(r) J_{\nu} \), where \( J_{\nu} \)
%is the angle-averaged intensity. 
The transfer equation with these
modifications is
\begin{eqnarray}
\mu \frac{\partial I_{\nu}}{\partial r} +
\frac{(1-\mu^{2})}{r} \frac{\partial I_{\nu}}{\partial \mu}
& = &
-\chi^{c}(r) I_{\nu}
+\kappa^{c}(r) B_{\nu}(T(r)) \nonumber \\
& + & \sigma^{c}(r) J_{\nu}
+ j_{\nu}(r)
\label{rt1}
\end{eqnarray}
We now integrate eq.(\ref{rt1}) over the 
spectral-line bandwidth appropriate to
either the H\(\alpha\) or the H\(\beta\) line. We assume that this
bandwidth is adequate to cover all the line radiation in the two
hydrogen lines studied, regardless of all Doppler shifts within the
source. In this connexion, we note that the velocity extent of the
line due to the bulk motion of expansion is of order 20\,km\,s\(^{-1}\),
%20\,\(v_{10}\)\,km\,s\(^{-1}\), where \(v_{10}\) is the terminal
%speed of the shell in units of 10\,km\,s\(^{-1}\), 
whilst the
thermal Doppler width is 21.4\,\(T_{4}^{1/2}\)\,km\,s\(^{-1}\),
with \(T_{4}\) equal to the kinetic temperature in the ionized
cavity in units of \(10^{4}\)\,K. An unknown microturbulent width
must be added in quadrature to the latter figure. It is therefore
reasonable to suppose that even the extreme red- and
blue-shifted portions of the line are significantly blended
by the thermal and microturbulent lineshape. In other words, the
combination of a low terminal expansion velocity and a large
thermal plus microturbulent line width allows us to neglect
velocity field induced Doppler shifts that would otherwise 
complicate the analysis considerably.
 Let the spectral-line bandwidth be \( \Delta \nu \), and the
line-integrated intensity is given by
\begin{equation}
I = \int_{-\Delta \nu/2}^{\Delta \nu/2} I_{\nu} d\nu
\label{filter}
\end{equation}
A similar equation to eq.(\ref{filter}) relates the angle-averaged
intensities \( J \) and \( J_{\nu} \). If we assume also that the
functions \( \chi^{c} \), \( \kappa^{c} \), \( B_{\nu}(T) \) and
\( \sigma^{c} \) vary only very slightly over \( \Delta \nu \),
these functions can be removed from the filter integral when it
is applied to eq.(\ref{rt1}). The result is
\begin{eqnarray}
\mu \frac{\partial I}{\partial r} +
\frac{(1-\mu^{2})}{r} \frac{\partial I}{\partial \mu}
& = &
-\chi^{c}(r) I
+\Delta \nu \kappa^{c}(r) B_{\nu}(T(r)) \nonumber \\
& + & \sigma^{c}(r) J
+ j(r),
\label{rt2}
\end{eqnarray}
where \( j(r) = \int_{-\Delta \nu/2}^{\Delta \nu/2} j_{\nu} d\nu \).
The only frequency-dependent part of \( j_{\nu} \) is the appropriate
lineshape function (see eq.(\ref{emalpha}) and eq.(\ref{embeta})).
We assume that the filter width is sufficient for the normalisation
condition of the lineshape to hold, so that the integral \( j(r) \)
in eq.(\ref{rt2}) is either,
\begin{equation}
j(H\alpha )=\frac{h\nu_{32}}{4\pi} A_{32} n_{3}(r)
\label{intalpha}
\end{equation}
or the equivalent expression for H\(\beta\).

Finally, we re-write eq.(\ref{rt2}) in separate forms appropriate
to the cavity, and to the shell, respectively. In the cavity, we
make the approximation that there is no dust, and that other 
processes, such as free-free and bound-free emission, make a
negligible contribution to the radiation flux within the 
filter bandwidths used, for example Matsuura et al. \shortcite{Matsuura05}.
%so all the continuum
%processes make no contribution: we ignore free-free emission since
%$hc/(\lambda k T) > 2$ at $T=10^4$\,K 
%for both the wavelengths studied. 
The cavity therefore acts 
effectively as a
pure source of line radiation, and the radiation transfer equation
in this zone is,
\begin{equation}
\mu \frac{\partial I}{\partial r} +
\frac{(1-\mu^{2})}{r} \frac{\partial I}{\partial \mu} = j(r).
\label{rtcav}
\end{equation}
By contrast, the shell, which is rich in dust, and
where almost all the hydrogen is molecular, does not emit any
H\(\alpha\) or H\(\beta\) line radiation, is a
 site of continuum absorbtion, scattering and thermal
emission by dust. The
shell transfer equation is therefore,
\begin{eqnarray}
\mu \frac{\partial I}{\partial r} +
\frac{(1-\mu^{2})}{r} \frac{\partial I}{\partial \mu}\! & = & \!
-\chi^{c}(r) I \nonumber \\
& + & \Delta \nu \kappa^{c}(r) B_{\nu}(T(r))
 +  \sigma^{c}(r) J
\label{rtshell}
\end{eqnarray}
We now proceed to solve eq.(\ref{rtcav}) and eq.(\ref{rtshell}) in
turn,
coupling the solutions via boundary conditions.
The solution is presented in several stages, and we outline these
briefly here. In Section~\ref{cav_sol}, we solve eq.(\ref{rtcav}) for
the specific intensity of line radiation in the cavity, and this
solution is angle-averaged in Section~\ref{cav_mean} yielding a mean intensity
in the cavity as a function of radius. This quantity is not required for
later calculations, so Section~\ref{cav_mean} is only included for
completeness. In Section~\ref{shell_sol}, we solve eq.(\ref{rtshell}) for
radiative transfer in the shell: initially the specific intensity
is obtained for dust extinction only, but a scattering source is then
re-introduced as a perturbation. The cavity and shell solutions, for the
specific intensity of a ray that may pass through both zones, are
combined in Section~\ref{comb_sol}. Section~\ref{secmean} is devoted to
developing a mean intensity in the shell, as a function of radius, by
angle-averaging the unperturbed solution from Section~\ref{comb_sol}.
This angle-averaged intensity is used to develop explicit forms for
the scattering perturbation in Section~\ref{sectpert}. The remaining sections
are concerned with observational properties of the solution, and
fits to HST data for NGC6537.

\section{The Cavity Solution}
\label{cav_sol}

\begin{figure}
\vspace{0.3cm}
\psfig{file=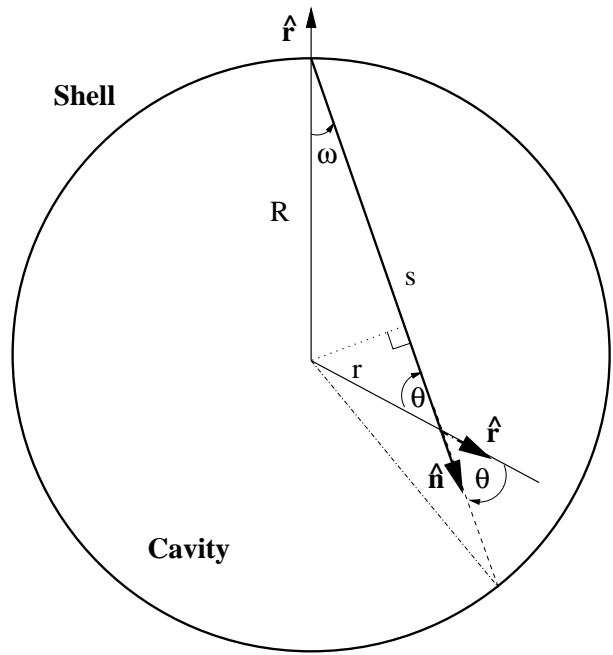,angle=0,width=8.0cm}
\caption{
A ray enters the cavity, of radius \( R \),
 at the top of the figure along path
\( s \) in direction \( \bmath{\hat{n}}\). The ray is initially
at an angle \( \omega \) to the radial unit vector,
\( \bmath{\hat{r}}\). This relationship is modified along
\( s \) as \( \bmath{\hat{r}}\) changes direction: at an
arbitrary point at radius \( r \) the ray unit vector
\( \bmath{\hat{n}}\) and the radial unit vector cross
at angle \( \theta \). Note that when \( s \) is smaller than
the distance to the mid-point of the chord, \( \theta \) 
becomes an obtuse angle.
}
\label{figcavity}
\end{figure}

The layout of a typical ray passing through the cavity zone
of the nebula is depicted in Fig.~\ref{figcavity}. The
model cavity has no inner radius. All rays
enter the cavity from the surrounding shell with some boundary
value of the specific intensity. This value may differ from ray
to ray, and, at present, we leave this value as an unknown.
The emission coefficient depends on the radius only through the
number density of H atoms in the upper level of the appropriate
Balmer line from eq.(\ref{intalpha}). We make the approximation
that these number densities are constants
in the cavity, on the grounds that the cavity gas is in pressure
equilibrium because of its high sound speed, and that we
have already assumed a constant temperature in this region. We
ignore the effect of overpressured regions near the edge of the
cavity where the ionized medium is juxtaposed against the
surrounding shell \cite{perinotto04}.
Therefore, we can re-cast eq.(\ref{rtcav}) as
\begin{equation}
\mu \frac{\partial I}{\partial r} +
\frac{(1-\mu^{2})}{r} \frac{\partial I}{\partial \mu} = j_{0},
\label{rtcav1}
\end{equation}
where \( j_{0} \) is the constant emission coefficient. The
left-hand side of eq.(\ref{rtcav1}) is the spherical polar
expansion of the dot product, \(\bmath{\hat{n}} \bullet
\bmath{\nabla} I\), but an alternative representation is
as the derivative, \( dI/ds\), taken along the ray characteristic.
An alternative way of writing eq.(\ref{rtcav1}) is therefore,
\begin{equation}
\frac{dI}{ds} = j_{0}
\label{rtcav2}
\end{equation}
which has the simple linear solution,
\begin{equation}
I(s) = I_{0} + j_{0} s,
\label{cavsoln}
\end{equation}
where \( I_{0}\) is the unknown intensity with which the ray
enters the cavity from the shell. The problem now consists only
of re-expressing the distance along the characteristic, \( s \),
in terms of the independent variables \( r \) and
\( \mu = \cos \theta \). To obtain such a relation, we apply
the sine rule to the triangle in Fig.~\ref{figcavity}, which is
bounded by \( r \), \( s \) and \( R \), yielding,
\begin{equation}
\frac{r}{\sin \omega} = \frac{R}{\sin \theta} =
\frac{s}{\sin (\omega + \theta)}.
\label{sinerule}
\end{equation}
Now both \( R \) and \( \omega \) are constants, so it follows from
the first equality in eq.(\ref{sinerule}) that
\begin{equation}
r \sin \theta = R \sin \omega = K,
\label{sconst}
\end{equation}
is a constant along the characteristic. 
Applying the cosine rule
to the same triangle yields a quadratic equation in \( s \). The
geometry of Fig.~\ref{figcavity} requires the positive root, so 
that the distance along the 
characteristic is,
\begin{equation}
s = r \cos \theta + (R^{2} - r^{2}\sin^{2} \theta )^{1/2}.
\label{sdist}
\end{equation}
Using the result in eq.(\ref{sconst}), and the definition of
\( \mu \), we obtain \( s = r\mu + (R^{2}-K^{2})^{1/2}\).
%Consideration of the ray entering the cavity, where \( s = 0 \)
%and \( 0 \leq \omega < \pi/2 \) requires the positive sign.
The constant expression under the square root reduces to
\( R\cos \omega \), so the characteristic distance is
\( s = r\mu + R\cos \omega \). The intensity along the
characteristic is therefore,
\begin{equation}
I(r,\mu) = I_{0} + j_{0} (r\mu + R\cos \omega ),
\label{cavsoln1}
\end{equation}
which is the required cavity solution for constant
emission coefficient.
It is trivial to prove, from eq.(\ref{cavsoln1})
that it satisfies eq.(\ref{rtcav1}).

\section{Mean Intensity in the Cavity}
\label{cav_mean}

\begin{figure}
\vspace{0.3cm}
\psfig{file=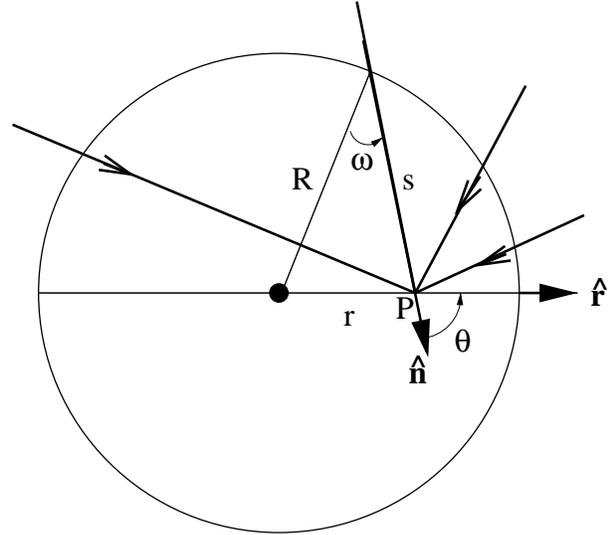,angle=0,width=8.0cm}
\caption{
The angle averaged intensity, \( J \), is to be computed
at point \( P \), at radius \( r \) from the centre. The average
is over rays (examples marked with arrows) which make angles,
\( \theta \), with the radius vector at \( P \), where
\( \theta \) may be in the range \( 0 \) to \( \pi \). A
particular ray is shown entering the cavity at angle
\( \omega \), in direction
\( \bmath{\hat{n}}\), and passing through a distance \( s \) in
order to reach \( P \).
}
\label{figjcav}
\end{figure}

To obtain the angle-averaged intensity, \( J(r) \), we
need to integrate over a number of different rays, which
all meet at one point, where the radius, \( r \), is a
constant. At this point, \( \theta \) is the final angle along
its path, and can be used in the required solid angle 
integral. However, now \( \omega \) is a variable, as we
are integrating over many rays, and \( \omega \) needs to
be expressed in terms of \( \theta \), or \( \mu \). The
situation is summarised in Fig.~\ref{figjcav}.
Results from the cavity solution, which still apply to
Fig.~\ref{figjcav}, are that \( \sin \omega =
(r/R)\sin \theta \) and \( s = r\cos \theta + R \cos \omega \).
We use the first of these relations to eliminate \( \omega \)
from the second. The expression for \( s \) can in turn be
used to write the cavity solution, eq.(\ref{cavsoln1}) in
the form,
\begin{equation}
I(r,\mu) = I_{0}(\mu) + j_{0} (r\mu + [(R^{2}-r^{2}) + r^{2}\mu^{2}]^{1/2},
\label{cavsoln3}
\end{equation}
noting that \( \mu = \cos \theta \) is the only variable on the
right-hand side of eq.(\ref{cavsoln3}).

\begin{figure}
\vspace{0.3cm}
\psfig{file=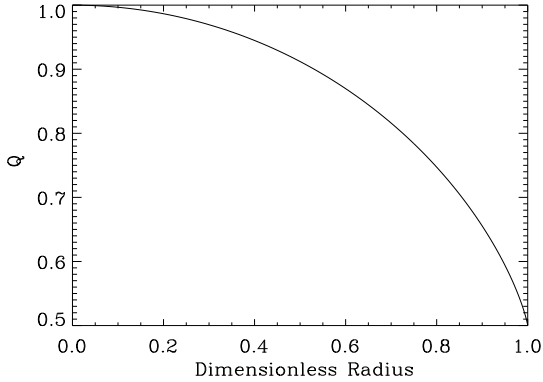,angle=0,width=8.0cm}
\caption{
A dimensionless form of the angle-averaged intensity in the
cavity, \( [J(r)-J_{0}]/(j_{0}R) \), see eq.(\ref{jbar2}), plotted
as a function of dimensionless radius, \( r/R \). 
Equation~\ref{jbarsmallr} is used for the point at \( r/R = 0 \).
}
\label{jcavfig}
\end{figure}

The solid angle integral for the mean intensity, with the integration
over the azimuthal angle already carried out is
\begin{equation}
J(r) = \frac{1}{2} \int_{-1}^{1} I(r,\mu ) d\mu .
\label{jdef}
\end{equation}
After substitution of eq.(\ref{cavsoln3}) into eq.(\ref{jdef}), and
a partial evaluation of the resulting integral, we find,
\begin{equation}
J(r) = J_{0} + \frac{1}{2} j_{0}r \int_{-1}^{1} (\alpha + \mu^{2})^{1/2}
d\mu ,
\label{jbar1}
\end{equation}
where \( \alpha = [(R/r)^{2}-1] \), and \( J_0 \) is the
angle average of \( I_0 \). The result is a standard
integral in terms of the arcsinh function, and after converting
this to logarithmic form, the angle-averaged intensity is
\begin{equation}
J(r) = J_{0} + \frac{j_{0}R}{2} \left[
1 + \frac{R}{r} \left( 1 - \frac{r^{2}}{R^{2}} \right)
\ln \left[
\frac{r+R}{\sqrt{R^{2}-r^{2}}}
    \right]
                                \right].
\label{jbar2}
\end{equation}
We note that the apparent infinity in eq.(\ref{jbar2})
at \( r=0 \) disappears when this equation is replaced by a
suitable expansion for the case \( r\ll R \). The small radius
form is,
\begin{equation}
J(r) \rightarrow J_{0} + (1/2) j_{0}R (2 - r^{2}/R^{2}).
\label{jbarsmallr}
\end{equation}
We plot, in Fig.~\ref{jcavfig}, the function
\( Q = (J(r)-J_{0})/(j_{0}R) \) from eq.(\ref{jbar2}), as a
function of the dimensionless radius \( x = r/R \).

\section{The Shell Solution}
\label{shell_sol}

We re-write the shell transfer equation as
\begin{equation}
\mu \frac{\partial I}{\partial r} +
\frac{(1-\mu^{2})}{r} \frac{\partial I}{\partial \mu}\! = \!
-\chi^{c}(r) I
 + f(r),
\label{rtshell1}
\end{equation}
where \( f(r) \) is an arbitrary function of the radius, which
incorporates the continuum emission and line scattering.
We change the left-hand side of eq.(\ref{rtshell1}), as in the
cavity solution, to describe a solution along the ray
characteristic. The equation can then be put in standard
first-order linear form as
\begin{equation}
\frac{dI}{ds} + \chi^{c}(r) I = f(r(s)).
\label{rtshell2}
\end{equation}
Equation~\ref{rtshell2} may be integrated by the standard method
of integrating factors. The boundary condition at \( s = 0 \) is
that the specific intensity enters the shell from interstellar
space with the intensity \( I_{BG} \), assumed equal to a typical
value for Galactic starlight. The extinction coefficient at this
same position is zero, regardless of its variation within the
shell. With this condition imposed, eq.(\ref{rtshell2}) has the
formal solution,
\begin{eqnarray}
I(s) & = & I_{BG} e^{-\int_{0}^{s}\chi^{c}(r(s'))ds'} \nonumber \\ & + &
\int_{0}^{s} f(r(s')) \exp \left\{
- \int_{s'}^{s}\chi^{c}(r(\sigma ))d\sigma
                           \right\} ds'.
\label{rtshell3}
\end{eqnarray}
The ray distance, \( s \), is related to the radius \( r \) and
direction cosine \( \mu \) by a modified form of the relation which
appears in eq.(\ref{cavsoln1}). If the shell has an outer radius \( R_{2} \),
then
\begin{equation}
s = r\mu + R_{2} \cos \omega,
\label{slen}
\end{equation}
where, for the moment, we ignore the presence of the cavity. With
this same caveat, \( r \sin \theta = R_{2} \sin \omega \) is a
constant along a given ray if \(\omega \) is the angle with which
the ray enters the shell from interstellar space.
Defining \( K_{2} = R_{2} \sin \omega \),
we can express \( \mu \) in terms of the radius as
\(
\mu = (1-K_{2}^{2}/r^{2})^{1/2}
\),
which may be substituted into eq.(\ref{slen}) to yield a relation
between \( r \) and \( s \) with no other variables involved. This
relation, with the radius as the subject, is
\begin{equation}
r = (s^{2}-2 s R_{2}\cos \omega + R_{2}^{2})^{1/2}.
\label{rdef}
\end{equation}
The expression for the radius in eq.(\ref{rdef}) can be used to
expand the radius in terms of \( s \) in eq.(\ref{rtshell3}), provided
that functional forms for \( \chi^{c}(r)\) and \( f(r) \) are
known.

\subsection{Shell Solution with Power-Law Density}

To progress beyond eq.(\ref{rtshell3}) we require some analytical
approximation for the behaviour of the radius-dependent functions.
We start with the extinction, \( \chi^{c} \). We assume that this
depends on the radius only through the number density of the
shell dust, so that the functional form is the same for both the
absorption and scattering contributions. Further, we assume a
density dependence which follows the inverse square behaviour of
the singular isothermal sphere. Therefore, we
suppose that the extinction in the
shell behaves as
\begin{equation}
\chi^{c}(r) = \chi^{c}(R) (r/R)^{-2},
\label{sissy}
\end{equation}
where \( \chi^{c}(R) \) is the maximum value of the extinction
coefficient, found just outside the cavity boundary. With the
help of eq.(\ref{rdef}), we can write the extinction as a function
of \( s \) rather than \( r \). Integrals of the type which
appear in eq.(\ref{rtshell3}) can now be written in the form,
\begin{equation}
\int \chi^{c}(r(s)) ds = R^{2}\chi^{c}(R)
\int \frac{ds}{s^{2}-2sR_{2}\cos \omega +R_{2}^{2}}.
\label{chiint}
\end{equation}
Introducing the new variables \( x = s-R_{2}\cos \omega \) and
\( a = R_{2} \sin \omega \), eq.(\ref{chiint}) may be written
as the standard integral,
\begin{equation}
\int \chi^{c}(r(s)) ds = R^{2}\chi^{c}(R)
\int \frac{dx}{x^{2}+a^{2}},
\label{chistd}
\end{equation}
which has the solution (as an indefinite integral),
\begin{equation}
\int \chi^{c}(r(s)) ds = \frac{R^{2}\chi^{c}(R)}{R_{2}\sin \omega}
\arctan \left[
              \frac{s-R_{2}\cos \omega}{R_{2}\sin \omega}
        \right].
\label{chisoln}
\end{equation}
Equation \ref{rtshell3} contains two definite forms of
eq.(\ref{chisoln}). The first of these has limits of \( 0 \)
to \( s \), and can be re-written, with the help of addition
formulae for the arctangent from Gradshteyn \& Ryzhik 
\shortcite{gr65}, as
\begin{equation}
\int_{0}^{s} \!\!\! \chi^{c}(r(s')) ds' = \frac{\beta}{\sin \omega}
\left[
\{\pi\} + \arctan \left(
   \frac{ s\sin \omega }{R_{2} - s\cos \omega }
                  \right)
\right],
\label{chisolA}
\end{equation}
where the \( \pi \) is to be included only for ray distances such
that \( s > R_{2} / \cos \omega \), and the group
\( \beta = R^{2}\chi^{c}(R)/R_{2} \).
The second definite integral form from eq.(\ref{rtshell3}) has the
limits \( s' \) to \( s \), and it is convenient to write it
without combining the arctangents, since the part with \( s \)
can be removed from the integral over \( s' \). Overall, the
shell solution can now be written,
\begin{eqnarray}
I(s) & \!\!\!\!\! = \!\!\!\!\! & I_{BG} \exp \left\{ \!
                   -\frac{\beta}{\sin \omega}
                 \left[ \{ \pi \} + \arctan \left( \!
                        \frac{\sin \omega}{(R_{2}/s)-\cos \omega}
                                    \!  \right)
                 \right]
              \!  \right\} \nonumber \\
     & \!\!\!\!\! + \!\!\!\!\! & e^{-\frac{\beta u(s)}{\sin \omega}}
     \int_{0}^{s} f(s') \exp \left\{
       \frac{\beta u(s')}{\sin \omega}
                             \right\} ds',
\label{rtshell4}
\end{eqnarray}
where
\begin{equation}
u(s) = \arctan \left[
   \frac{s-R_{2}\cos \omega}{R_{2}\sin \omega}
               \right]
\label{udef}
\end{equation}
and the functional form of \( f \) remains to be determined.

\subsection{Scattering as a perturbation}
\label{pertstuff}

If we ignore continuum emission in the shell, the unknown
function, \( f \) in eq.(\ref{rtshell4}) reduces to line
scattering alone, which depends on the angle-averaged mean
intensity, \( J(r) \). It is then possible, in principle, to
average eq.(\ref{rtshell4}) over solid angle, leading to an
integro-differential equation for \( J(r) \). However, owing
to the difficulty in attempting to solve such an equation
analytically, we resort to the simpler procedure of treating
the scattering term as a perturbation. For this approximation
to be very good, the scattering contribution to the extinction
should be small, so that \( \sigma^{c}(r)/\chi^{c}(r)\) is a
small parameter for all radii. This is unlikely to be true
for real dust. For example, silicate dust modelled by Ossenkopf,
Henning \& Mathis \shortcite{ossenkopf92}
has an optical efficiency for scattering which is consistently
\(\sim\)2-3 times that for absorption over optical wavelengths.
However, we still adopt the perturbative approach as the
only viable method of obtaining an approximate analytical
solution.
In this procedure, we first solve the radiative
transfer equation in the shell, asuming that line extinction
is the only contribution to the right-hand side of
eq.(\ref{rtshell1}). This means that we can set \( f = 0 \)
in eq.(\ref{rtshell4}), and take as the zero-order solution
in the shell,
\begin{eqnarray}
I(r,\mu) \!\!\!\!\!\! & = & \!\!\!\!\! I_{BG} \exp \left\{
   \frac{-\beta R_{2}}{r(1\!-\!\mu^{2})^{1/2}} \left[
    (\pi )
                                           \right.
                            \right. \nonumber \\
     \!\!\!\!\!\!\!\! & + & \!\!\!\!\!\!\!\!
                                       \left.    \left.
   \arctan \! \left( \!\!
      \frac{(1\!-\!\mu^{2})^{1/2}[r\mu \! +
 \!(R_{2}^{2}\!-\!r^{2}(1\!-\!\mu^{2}))^{1/2}]}
           {r(1\!-\!\mu^{2}) -\mu (R_{2}^{2}\!-\!r^{2}(1\!-\!\mu^{2}))^{1/2}}
           \! \right)
                                        \! \!  \right]
                        \!  \!  \right\},
\label{rtshell5}
\end{eqnarray}
which is just the first term of eq.(\ref{rtshell4}), with
\( \theta \) and \( \omega \) eliminated in favour of
\( \mu = \cos \theta \), and with the help of the relation
\( \sin \omega = (r/R_{2}) \sin \theta \) (see eq.(\ref{sconst}),
which still holds in the shell). It is tedious, but straightforward,
to show that eq.(\ref{rtshell5}) is indeed a solution
of eq.(\ref{rtshell1}), for the case where \(f(r)=0\).
Equation~\ref{rtshell5} cannot
be used alone, as along any path through the nebula, a ray
may encounter a sequence of shell, cavity, and again shell
conditions. We therefore need to determine some functional
form for \( I_{BG} \) for the case where a ray leaves the
cavity and re-enters the shell, forming a combined solution
along a ray.

As scattering is now to be treated as a perturbation, it can
be computed from a mean intensity which is derived from the
zeroth-order combined solution. An equation for a perturbation
in the intensity, \( \delta I \), can be constructed from
eq.(\ref{rtshell}) by expanding the specific intensity
as \( I = I_{*} + \delta I \), and then subtracting off the
equation in the zeroth-order estimate, \( I_{*} \). The result
is,
\begin{equation}
\mu \frac{\partial (\delta I)}{\partial r} +
\frac{(1-\mu^{2})}{r} \frac{\partial (\delta I)}{\partial \mu}\! = \!
-\chi^{c}(r) \delta I
 + \sigma^{c} J(r),
\label{rtpert}
\end{equation}
where we have assumed that continuum emission makes a negligible
contribution, so that \( f(r) = \sigma^{c} J(r) \). So,
mathematically, the equation in the perturbation may be
treated in the same way as the shell equation,
eq.(\ref{rtshell1}). The formal solution for \( \delta I \)
therefore looks like eq.(\ref{rtshell3}) with the scattering
term replacing the unknown function \( f \). 
% For mathematical
%convenience, we draw a distinction between the two terms on the
%right-hand side of eq.(\ref{rtpert}): the scattering term is
%zeroth order in the radiation variables, whilst the extinction
%term is first order. We ignore the extinction term, even though
%the scattering term is also formally first order because we
%are assuming \( \sigma^{c} \ll \chi^{c} \). This rather dubious
%assumption reduces eq.(\ref{rtpert}) to
The perturbation solution is therefore given by
%\begin{equation}
%\mu \frac{\partial (\delta I)}{\partial r} +
%\frac{(1-\mu^{2})}{r} \frac{\partial (\delta I)}{\partial \mu}\! = \!
% + \sigma^{c} J(r).
%\label{rtpert2}
%\end{equation}
\begin{eqnarray}
\delta I(s) & = & \delta I(s_{0}) e^{-h(s)}
 \nonumber \\ & + &
e^{-h(s)}\int_{s_{0}}^{s} \sigma^{c}(r(s')) J(r(s')) 
e^{h(s')} ds',
\label{rtpert2}
\end{eqnarray}
where the function \( h \) is given by eq.(\ref{chisoln}).
%An important point about eq.(\ref{rtpert2}) is that it is
%mathematically
%the same as the cavity equation, with the exception that the
%right-hand side is a non-constant function of radius.
It is important to note that there are three possible versions
of eq.(\ref{rtpert2}): A ray which avoids the cavity has
\( \delta I(s_{0}) = 0 \) and a lower limit of \( s_{0} = 0 \)
on the remaining integral. A ray which enters the cavity has
two shell segments. In the first, \( \delta I(s_{0}) = 0 \) and
\( s_{0} = 0 \), but the upper limit is the value of the
path where the ray enters the cavity. The perturbation is then
constant in the cavity, forming a finite value of
\( I(s_{0}) \) for the second shell segment, where \( s_{0} \)
is now the path length where the ray leaves the cavity.

\section{The Combined Solution Along a Ray}
\label{comb_sol}

\begin{figure}
\vspace{0.3cm}
\psfig{file=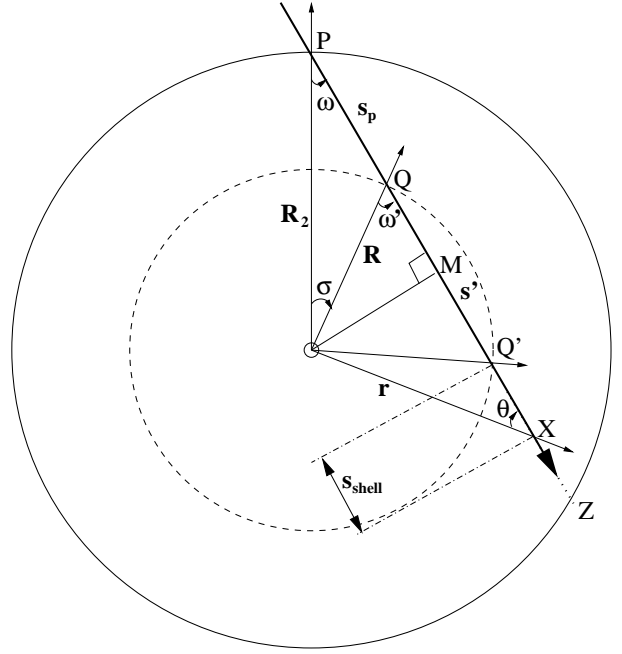,angle=0,width=8.0cm}
\caption{
The complete ray path is shown divided into three segments: between
P and Q a distance \( s_{p} \) is covered in the shell, having
entry angle \( \omega \). The ray
enters the cavity at Q, with angle \( \omega ' \),
and covers the distance \( s' \) between
Q and Q' before exiting into the shell. The point M is the mid-point
of \( s' \), and the closest approach of the ray
to the central star (at O). Once in the shell again, the ray reaches
the general point X, at radius \( r \), where its path makes an
angle, \( \theta \), with the radial direction. The ray direction
vector and radial direction vector (at various positions) are
marked with bold and light arrows respectively. In progressing from
Q' to X, the ray covers the distance \( s_{shell} \). The ray
finally leaves the nebula at point Z, where \( r = R_{2} \).
}
\label{figcomb}
\end{figure}

Given the approximations made in the previous section, it is
now possible to combine the cavity and shell solutions for a
single ray. The geometry of the combined solution is set out
in Fig.~\ref{figcomb}. As we are ignoring, as this stage, rays
which pass solely through the shell, each effective ray has
three segments: the first is an absorptive passage through the
shell. However, if the input background at \( r=R_{2} \) is very
weak, we can ignore this segment, and treat the specific intensity
of the ray at the cavity boundary, \( r = R \) as zero.

The second segment of the ray path, between points Q and Q' in
Fig.~\ref{figcomb}, passes through the cavity. Here the specific
intensity is assumed to increase through spontaneous emission in
the line, according to the cavity solution, eq.(\ref{cavsoln1}). In
particular,
we want an expression for the specific intensity at Q', where
the cavity solution becomes the input value for the shell
solution, which takes over as the third segment of the ray
path, which continues until point Z is reached, and the ray
passes into the vacuum.

At the point X in Fig.~\ref{figcomb}, which is
as radius \( r \), the ray has travelled a
distance \( s_{shell} \) through the shell, and it makes an
angle \( \theta \) with the radial vector. At this point, the
solution will be the shell solution, with an input specific
intensity from the cavity.

Analysis of the triangle MOQ shows that the cavity segment
of the ray path has length \( s' = 2R\cos \omega ' \). The
total distance from point P to point M is, from the triangle
MOP, equal to \( s_{M} = R_{2}\cos \omega \). Therefore, the
`pre-cavity' distance (from P to Q) is
\( s_{p} = R_{2}\cos \omega - R \cos \omega ' \).
 The additional
distance along the ray from M to X is \( s_{M'} = r\cos \theta
= r\mu \). The total distance along the ray from P to X
is therefore \( s = R_{2} \cos \omega + r\mu \), just as in
the shell solution, eq.(\ref{slen}). From these results,
we find that the distance \( s_{shell} \) in
Fig.~\ref{figcomb} is given by
\begin{equation}
s_{shell} = r\mu - R \cos \omega ',
\label{sshell}
\end{equation}
and by applying the sine rule to the triangle OPQ, the
angles \( \omega '\) and \( \omega \) are related by
\begin{equation}
\sin \omega ' = (R_{2}/R) \sin \omega,
\label{omegag}
\end{equation}
which can be used to eliminate \( \omega ' \) from eq.(\ref{sshell}),
yielding
\begin{equation}
s_{shell} = r\mu - [R^{2} - R_{2}^{2} \sin^{2} \omega ]^{1/2}.
\label{sshell2}
\end{equation}
The limiting (maximum) value of \( \omega \), for a ray which 
just enters the cavity, occurs when \( \omega ' = \pi / 2 \),
when \( \sin \omega_{max} = R/R_{2} \).

\subsection{Input Solution to the Shell}
\label{inputshell}

As we are assuming that the specific intensity at point Q
in Fig.~\ref{figcomb} is effectively zero, the input solution
for the shell is just the cavity solution evaluated at Q'.
%The situation of Fig.~\ref{figcavity} can be recovered by
%rotating Fig.~\ref{figcomb} anticlockwise through the
%angle \( \sigma \). 
When this is done, the cavity solution,
eq.(\ref{cavsoln1}) applies, with \( \omega ' \) replacing 
\( \omega \). We take this solution with \( I_{0} = 0 \),
\( r = R \) and \( \mu = \cos \omega ' \), obtaining
\( I(R,\omega ') = 2j_{0}R\cos \omega' \). With the help
of eq.(\ref{omegag}), this expression becomes
\begin{equation}
I(R,\omega ) = 2j_{0}[R^{2} - R_{2}^{2} \sin^{2} \omega ]^{1/2},
\label{shellbg}
\end{equation}
which is the background for the solution in the shell
beyond point Q'.

\subsection{Combined Solution in the Shell}
\label{combinedinshell}

\begin{figure}
\vspace{0.3cm}
\psfig{file=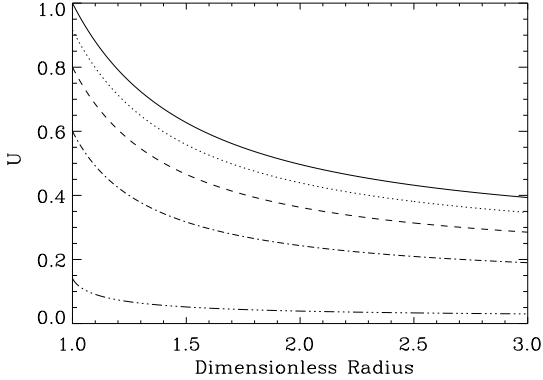,angle=0,width=8.0cm}
\caption{
The function \( U = I(x,a )/(2j_{0}R) \) (see eq.(\ref{combplot})
plotted as a function of dimensionless radius \( x = r/R \) for
values of \( a = 0.0\) (solid line), 0.4
(dotted line), 0.6 (dashed line), 0.8 (simple chain), 0.99
(complex chain).
}
\label{specplot}
\end{figure}

As the limits of integration are modified for the combined
solution, we generate the shell solution from the indefinite
integral, eq.(\ref{chisoln}). The upper limit is the distance
along the ray path, \(s \), but the lower limit is now, from
Fig.~\ref{figcomb}, \( s' + s_{p} = R_{2}\cos \omega
+ [R^{2} - R_{2}^{2}\sin^{2} \omega ]^{1/2}\). The result for
the extinction integral is
\begin{eqnarray}
\int_{s'+s_{p}}^{s} \chi^{c}(x)dx &  = &
\frac{R^{2}\chi^{c}(R)}{R_{2}\sin \omega} \left\{
\arctan \left[
   \frac{s-R_{2}\cos \omega}{R_{2}\sin \omega}
        \right] \right. \nonumber \\
 & - &                                   \left.
 \arctan \left[
   \frac{(R^{2}-R_{2}^{2}\sin^{2}\omega )^{1/2}}{R_{2}\sin \omega}
               \right]
                                          \right\}.
\label{shellext}
\end{eqnarray}
When eq.(\ref{slen}) has been applied, and eq.(\ref{shellext}) has
been re-expressed entirely in terms of \( \mu \), the result is,
\begin{eqnarray}
\int_{s'+s_{p}}^{s} \chi^{c}(x)dx \!\!\!\! &  = & \!\!\!\!
\frac{\beta}{(1-\mu^{2})^{1/2}} \left\{
\arctan \left[
   \frac{\mu}{(1-\mu^{2})^{1/2}}
        \right] \right. \nonumber \\
\!\!\!\!  & - & \!\!\!\!                            \left.
 \arctan \left[
   \frac{(1-(r/R)^{2}(1-\mu^{2}) )^{1/2}}{(r/R)(1-\mu^{2})}
               \right]
                                          \right\}.
\label{shellext2}
\end{eqnarray}
%where \( \beta = R^{2}\chi^{c}(R)/R_{2} \) is a constant. 
The first
arctangent can be converted to a simpler form, as an arcsine, via
the relation
\begin{equation}
\arctan \left[
   \frac{x}{(1-x^{2})^{1/2}}
        \right] = \arcsin x,
\label{gr1}
\end{equation}
which is taken from Gradshteyn \& Ryzhik \shortcite{gr65}. The second
arctangent in eq.(\ref{shellext2}) can be written as the reciprocal of
the form which appears in eq.(\ref{gr1}) by defining the
new variable \( q = r(1-\mu^{2})/R \). An additional relation
linking arctangents \cite{gr65} for the case where \( x > 0 \), which
is true of \( q \), is
\begin{equation}
\arctan (1/x) = \pi / 2 - \arctan x,
\label{gr2}
\end{equation}
and this relation allows us to use eq.(\ref{gr1}) directly on the
second arctangent in eq.(\ref{shellext2}). A third relation from
Gradshteyn \& Ryzhik, \( \arcsin x + \arccos x = \pi / 2 \) allows us
to re-write eq.(\ref{shellext2}) as,
\begin{equation}
\int_{s'+s_{p}}^{s} \chi^{c}(x)dx = 
\frac{\beta}{(1-\mu^{2})^{1/2}} \left[
   \arcsin q - \theta
                                \right].
\label{shellext3}
\end{equation}
It is now a simple matter to write out the combined solution for a
ray by inserting eq.(\ref{shellext3}) into the shell solution,
eq.(\ref{rtshell3}), with \( I_{BG} \) given by eq.(\ref{shellbg}). With
\( \beta \) and \( q \) fully expanded in terms of \( \theta \), the
result is 
\begin{eqnarray}
I(r,\theta )& = & 2j_{0}R
            [1-(R_{2}/R)^{2}\sin^{2} \omega ]^{1/2} \nonumber \\
         &\times &    \exp \left\{
   \frac{-R\chi^{c}(R)}{(r/R)\sin \theta} \left[
       \arcsin \left(
          \frac{r\sin \theta}{R}
               \right) - \theta
                                          \right]
                       \right\},
\label{combsoln}
\end{eqnarray}
which is valid for \( R \leq r \leq R_{2} \). For plotting, we
re-write eq.(\ref{combsoln}) in the form
\begin{equation}
I(x,a) = 2j_{0}R (1-a^{2})^{1/2} \!
              \exp \!\! \left[ -\frac{\tau}{a}(\arcsin a - \arcsin \frac{a}{x})
                   \right],
\label{combplot}
\end{equation}
an equation in two parameters, \( a = (R_{2}/R)\sin \omega \), and the
optical depth parameter,
\( \tau = R\chi^{c}(R) \). The latter parameter becomes equal to
the radial optical depth of the shell in the limit where \( R_2 \gg R \). 
The independent variable is \( x = r/R \). This last
definition implicitly introduces a third parameter, the upper bound
on \(x\), equal to \( R_{2}/R \). We note that eq.(\ref{combplot})
reduces to a sensible limiting form, that is
\begin{equation}
I(x,a=0) = 2j_0 R \exp [ -\tau (1-R/r) ],
\label{eq_thelastone}
\end{equation}
for the case where \( a = 0 \).
In Fig.~\ref{specplot} we show the function \( U = I(x,a )/
(2j_{0}R) \) for five values of \( a \) between its minimum
value of zero, and maximum of 1. We set the optical depth parameter to
be \( \tau = 1.38 \) (see 
Section~\ref{txt_parms}) to agree approximately with the visual extinction
of \(1.5\)\,magnitudes near the centre of NGC6537, and we plot the
dimensionless radius out to \( x = 3 \).

\section{Mean Intensity in the Shell}
\label{secmean}

The approximation discussed in Section~\ref{inputshell} - that rays
that do not cross the cavity contribute zero specific intensity - excludes
all rays with negative values of $\cos \theta$. Rays with non-zero
specific intensity are limited to a subset of those with positive
values of $\cos \theta$, more precisely to 
those with \( \sin \theta < R/r \). With the additional restriction
that any radius where \( J(r) \) is computed
has \( R \leq r \leq R_{2} \), the
situation here is similar to that shown in Fig.~\ref{figjcav}.
The function to be
averaged is the combined solution specific intensity in the
zeroth-order (no scattering) approximation, given by
eq.(\ref{combsoln}). The above considerations allow us to write
a formal integral for the mean intensity, recalling that
\( r \sin \theta = R_2 \sin \omega \):
\begin{eqnarray}
J(r)\!\!\!\! &  = & \!\!\!\! j_{0} R \int_{0}^{\arcsin (R/r)}
        [1-(r/R)^{2}\sin^{2}\theta ]^{1/2} \nonumber \\
\!\!\!\!    & \times & \!\!\!\!  \exp \left\{ \!
      \frac{R\chi^{c}(R)}{(r/R)\sin \theta}
         \left[
            \arcsin \left(
               \frac{r\sin \theta}{R}
                    \right) -\theta
         \right]
                    \!  \right\} \sin \theta d\theta
\label{jbarshell}
\end{eqnarray}

We simplify eq.(\ref{jbarshell}) via a series of substitutions.
The first of these is to let the impact parameter be
\( p = (r/R)\sin \theta \); we
also define the new constant parameters, \( \gamma = R\chi^{c}(R) \),
and \( \rho = R/r \). These definitions
transform eq.(\ref{jbarshell}) to,
\begin{eqnarray}
J(r) & = & j_{0} R \rho^{2} \int_{0}^{1} p
       \left(
          \frac{1-p^{2}}{1-\rho^{2}p^{2}}
       \right)^{1/2} \nonumber \\
     & \times & \exp \left\{
         \frac{-\gamma}{p} \left[
             \arcsin p - \arcsin \left(
               \rho p
                                 \right)
                           \right]
                     \right\} dp,
\label{jbarshell2}
\end{eqnarray}
noting  that \( p \) and \( \rho \) are always \( \leq 1 \).
Given this condition, we can expand the inverse sines in
eq.(\ref{jbarshell2}) in terms of Gauss hypergeometric
functions (for example, Abramowitz \& Stegun \shortcite{absteg65}).
For complex argument \( z \),
\begin{equation}
\arcsin z = z F(1/2,1/2;3/2,z^{2}),
\label{gauss}
\end{equation}
where the power series forming the Gauss hypergeometric function,
\( F \), is absolutely convergent within, and on, the unit
circle for the arguments in eq.(\ref{gauss}) 
\cite{gr65}. The substitution of eq.(\ref{gauss}) into
eq.(\ref{jbarshell2}) has the beneficial consequence of cancelling
\( p \) within the exponential, and leaving \( p^{2} \) everywhere
except for the product \( p dp \). This suggests the substitution
\( x = p^{2} \), yielding,
\begin{eqnarray}
J(\rho )\!\!\!\!\! & = & \!\!\!\!\! \frac{j_{0}R\rho^{2}}{2} \int_{0}^{1}
           \left(
             \frac{1-x}{1-\rho^{2}x}
           \right)^{1/2} \nonumber \\
  \!\!\!\!\!       & \times & \!\!\!\!\! \exp \left\{ \!\!
            -\gamma \! \left[
   F\! \left(\frac{1}{2},\frac{1}{2};\frac{3}{2},x\right) \! - \!
\rho F \! \left(\frac{1}{2},\frac{1}{2};\frac{3}{2},\rho^{2}x\right)
                 \!   \right]
                      \!   \right\} dx
\label{jbarshell3}
\end{eqnarray}
At this point, we expand the hypergeometric series which appear in
eq.(\ref{jbarshell3}). For the case here, where the first two
arguments are the same, the power series \cite{absteg65} is
\begin{eqnarray}
F(a, a ; g,z)& = & 1 + \frac{a^{2}z}{g 1!} + 
                    \frac{a^{2}(a+1)^{2}z^{2}}{g(g+1)2!}
                    \nonumber \\
             & + &
                    \frac{a^{2}(a+1)^{2}(a+2)^{2}z^{3}}{g(g+1)(g+2)
                    3!}
               + ....,
\label{hyper}
\end{eqnarray}
in the case of general \( a \), \( g \), and for the specific 
case of \( a=1/2 \) and \( g = 3/2 \), we find,
\begin{equation}
F(1/2, 1/2 ; 3/2,z) = 1 + \frac{z}{6} + \frac{3z^{2}}{40}
                      + \frac{5z^{3}}{112} +...
\label{hyper2}
\end{equation}
Substitution of eq.(\ref{hyper2}) in eq.(\ref{jbarshell3}) leads
to the expression
\begin{eqnarray}
J(\rho )\!\!\!\! & = & \!\!\!\! \frac{j_{0}R\rho^{2}}{2} \int_{0}^{1}
           \left(
             \frac{1-x}{1-\rho^{2}x}
           \right)^{1/2} \exp \left\{ -\gamma
                                 \left[
              (1-\rho )
                                 \right. 
                              \right.  \nonumber \\
 \!\!\!\!\!\!       & + & \!\!\!\!\!\!
                                 \left.
                              \left.
              \frac{(1-\rho^{3})x}{6}
           +  \frac{3(1-\rho^{5})x^{2}}{40}
           + \frac{5(1-\rho^{7})x^{3}}{112}
           + ...
                                 \right] \!\!
                              \right\},
\label{jbarshell4}
\end{eqnarray}
which is as far as it is possible to proceed without making
further approximations.

\subsection{Approximate Forms}

Although the series in eq.(\ref{jbarshell4}) is convergent for
all relevant values of \( x =[(r/R)\sin \theta ]^{2} \), 
convergence is rather slow when
\( x \) is close to \( 1 \), the
condition for a ray in glancing contact
with the cavity. However, the contribution of
large \( x \) to the integral is limited by the term in front
of the exponential which tends to zero as \( x \rightarrow 1 \)
(for the particular case of \( \rho = 1 \), see below).
The physical reason for this is that rays at large angle have
relatively short paths through the cavity, and correspondingly
small specific intensities on entry to the shell.

Although values of \( \rho \) which are close to \( 1 \) (positions
close to the shell/cavity boundary) can force the leading term
in eq.(\ref{jbarshell4}) to \( 1 \), this is not a problem
because, in this situation, the argument of the exponential
approaches zero, the whole exponential to \( 1 \), and integration
of the leading term alone is sufficient for moderate accuracy.
We evaluate eq.(\ref{jbarshell4}) to two levels of accuracy in
\( x \). In the zero-order approximation, we abandon all but the
first term in the series, but this does not depend on \( x \), and
can therefore be removed from the integral, leaving,
\begin{equation}
J(\rho ) \simeq \frac{j_{0}R\rho^{2}}{2} e^{-\gamma (1-\rho )}
         \int_{0}^{1} \left(
   \frac{1-x}{1-\rho^{2}x}
                      \right)^{1/2} dx.
\label{jzero}
\end{equation}
On evaluation of the integral in eq.(\ref{jzero}), and reverting
to the original radius variable, \( r = R /\rho \), we obtain
the zero-order approximation to the mean intensity:
\begin{equation}
J(r) = \frac{j_{0}R}{2} e^{-\gamma (1-R/r)}
        \left[
           1 - \frac{r^{2}-R^{2}}{2rR} \ln \left(
               \frac{r+R}{r-R}
                                           \right)
        \right] .
\label{jbarzero}
\end{equation}
We note that at the cavity/shell boundary, where \( r = R \), the
reduction of eq.(\ref{jbarzero}) agrees with the cavity solution,
eq.(\ref{jbar2}), evaluated at the same radius. Both equations
yield \( J(R) = j_{0} R/2 \).

It is also possible to obtain an analytic integral for a first
order approximation, keeping the term in \( x \) in the series
in eq.(\ref{jbarshell4}). The integral in the first-order case
is a standard form in Gradshteyn \& Ryzhik \shortcite{gr65}.
With the parameters specific to the current problem, the
integral is
\begin{eqnarray}
\int_{0}^{1} \!\!\! \left( \!\!
   \frac{1-x}{1-\rho^{2}x} \!\!
                      \right)^{1/2} \!\!\!\!\!\!
   \exp \! \left\{\!\frac{-\gamma (1-\rho^{3})x}{6} \! \right\}\! dx 
\!\!\!\!\! &  = & \!\!\!\!\! B(1,\frac{3}{2})
  \nonumber \\ \!\!\!\! & \times & \!\!\!\!\!\!
  \Phi_{1} (1,\!\frac{1}{2},\!\frac{5}{2};\rho^{2},z ),
\label{nasty}
\end{eqnarray}
where \( z = -\gamma (1-\rho^{3})/6 \). The first function, \( B \),
on the right-hand side of eq.(\ref{nasty}) is an Euler beta-function,
whilst the second is a 
confluent hypergeometric series in two variables,
\( \rho^{2} \) and \( z \). Further standard formulae
allow the beta-function \( B(1,3/2) \) to be expressed as a 
ratio of factorials that reduces to \( 2/3 \), so that the
mean intensity to first order in \( x \) is
\begin{equation}
J(\rho ) = \frac{j_{0}R\rho^{2}}{3} e^{-\gamma (1-\rho )}
           \Phi_{1} \left(
          1,\frac{1}{2},\frac{5}{2};\rho^{2},\frac{-\gamma (1-\rho^{3})}{6}
                    \right)
\label{jbarone}
\end{equation}
\begin{figure}
\vspace{0.3cm}
\psfig{file=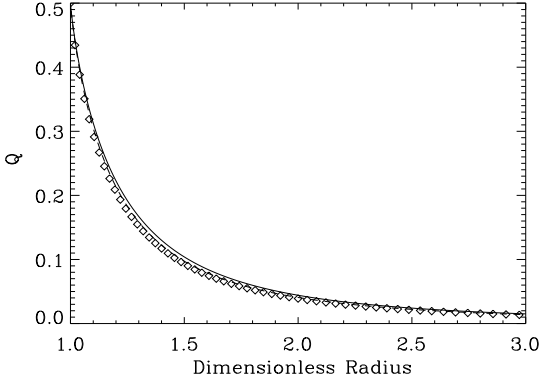,angle=0,width=8.0cm}
\caption{
The function \( Q = J(x)/(j_{0}R) \)
plotted as a function of dimensionless radius \( x = 1/\rho =r/R \)
using
the zeroth-order approximation, eq.(\ref{jbarzero}),
 (solid line), the first order approximation, eq.(\ref{jbarone}),
(dashed line), and exact integrals from eq.(\ref{jbarshell2})
(diamond symbols).
}
\label{jshellplot}
\end{figure}

We plot, in Fig.~\ref{jshellplot}, the mean intensity in the shell
as a function of dimensionless radius, \(
1/\rho \), computed from the exact
integral, eq.(\ref{jbarshell2}), and both approximations
(eq.(\ref{jbarzero}) and eq.(\ref{jbarone})). For the value of
the optical depth parameter used ($\gamma$ = 1.38, see
 Section~\ref{combinedinshell}),
we see that even the zeroth-order formula is an excellent
approximation. The graphs in Fig.~\ref{jshellplot} should be
seen as an extension of that in Fig.~\ref{jcavfig} to values of
\( r/R > 1 \) and to \( Q < 0.5 \) in the case where the
input background is negligible.

\section{Perturbation Solution in the Shell}
\label{sectpert}

We now proceed to solve eq.(\ref{rtpert2}) with the mean intensity
in the scattering term given by eq.(\ref{jbarzero}): it is too
complicated to use the more accurate eq.(\ref{jbarone}). There are
various forms of eq.(\ref{rtpert2}) which should be used for the
appropriate zone of the source. For a ray which penetrates the
cavity, we initially consider the case where this ray has entered
the shell, but has not yet reached the cavity. Assuming a negligible
input intensity from the vacuum, we have the boundary condition
that \( s_{0} = 0 \) and \( \delta I (0) = 0 \), and therefore the
problem for this ray, and zone, reduces to solving the integral,
\begin{equation}
\delta I(s) = 
e^{-h(s)}\int_{0}^{s} \sigma^{c}(r(s')) J(r(s')) 
e^{h(s')} ds'.
\label{ray1shell1}
\end{equation}
The most useful form for the function \( h(s') \), which
appears in the integrating factor, 
for this zone is
\begin{equation}
h(s') = \frac{R^{2} \chi^{c}(R) [\pi -\omega - \theta (s')]}
             {R_{2} \sin \omega},
\label{hshell1}
\end{equation}
which is derived from eq.(\ref{chisolA}), with a lower limit of
\( 0 \) and an upper limit of \( s' \), followed by transformations
similar to those used in working from eq.(\ref{gr1}) 
to eq.(\ref{combsoln}). Note in particular that for a ray in this
case, $\omega$ is an acute angle, but $\theta$ is obtuse, and the
respective limits of these angles for a radial ray are zero and $\pi$.

We substitute for the scattering coefficient, \( \sigma^{c}(r(s')) \),
by assuming that it has the same \( 1/r^{2} \) functional dependence
as the absorption (see Section~5.1). The mean intensity is given
by eq.(\ref{jbarzero}). It is perfectly possible to express all these
functions in terms of the distance, \( s' \), along the ray, but
the integral in eq.(\ref{ray1shell1}) appears easier when working
in terms of radius. Letting the integral be \( \Psi \), we have
\begin{eqnarray}
\Psi \!\!\!\!\! & = & \!\!\!\! \frac{\sigma^{c}(R) R^{2}j_{0} 
\exp \left[ \epsilon (\pi -\omega -
    a)\right]}{2}
\nonumber \\ \!\!\!\!\!\!\! & \times & \!\!\!\!\!\!\!\!\!
\int_{P_{0}}^{P} \!\!\! \frac{d\rho \exp 
\left[-\epsilon (\arcsin a\rho - a\rho)\right]}
                       { (1 - a^{2} \rho^{2} )^{1/2} }
 \!\! \left[ \! 1 \! - \! \frac{(1-\rho^{2})}{2\rho}
            \! \ln \!\!\! \left(
               \!\! \frac{1+\rho}{1-\rho} \!\!
                 \right) \!\!
  \right],
\label{omeq1}
\end{eqnarray}
where \( \rho = R/r \) as before, \( a = (R_{2}/R)\sin \omega \),
\( \epsilon = \gamma / a \), and the limits are given by
\begin{equation}
P_{0} = R/R_{2}
\label{p1}
\end{equation}
and 
\begin{equation}
P = R/[s^{2} - 2sR\cos \omega + R_{2}^{2}]^{1/2},
\label{p2}
\end{equation}
which reduces to \( 1 \) for a ray entering the shell from the
vacuum, and reaching the edge of the cavity, where \( r = R \). 
%Note
%that the running variable \( p \) has a different definition from
%the \( p \) used in Section~\ref{secmean}.
If we divide eq.(\ref{omeq1}) by the top line, which is independent
of \( \rho \), we can carry out an integration by parts. The integrated
part is 
\begin{equation}
V = \int \frac{\exp \left[-\epsilon \arcsin (a\rho)\right]}
              {(1-a^{2}\rho^{2})^{1/2}} d\rho  =
   -\frac{\exp \left[-\epsilon \arcsin (a\rho)\right]}
         {\epsilon a},
\label{vpart}
\end{equation}
which removes the problematic square-root term. The result of
this integration is,
\begin{eqnarray}
\Psi_{j} \!\!\!\! & = & \!\!\!\! - 2 \left[
             e^{-\epsilon (a-a\rho+\arcsin (a\rho))} \left(
   1 - \frac{(1-\rho^{2})}{2\rho} \ln \frac{1+\rho}{1-\rho}
                                            \right)
                \right]_{R/R_{2}}^{P} \nonumber \\
\!\!\!\! & + & \!\!\!\! \int_{R/R_{2}}^{P} g(\rho) \exp \left\{ -\epsilon
          \left(
            a - a\rho + \arcsin (a\rho)
          \right)
                                         \right\} d\rho,   
\label{omeq2}
\end{eqnarray}
where the function \( g(\rho) \), now entirely composed of
logarithms and rational functions of \( \rho \), is defined by
\begin{equation}
g(\rho) = 2\epsilon a -    2   \frac{1}{\rho} +
       \left[
         1 - \frac{\epsilon a}{\rho} + \frac{1}{\rho^{2}} + \epsilon a \rho
       \right] \ln \left(
             \frac{1+\rho}{1-\rho}
                   \right)
\label{gdef}
\end{equation}
and \( \Psi_{j} \) is given by
\begin{equation}
\Psi_{j} = \frac{4 \epsilon a \Psi} {\sigma^{c} (R) R^{2} j_{0}
             e^{\epsilon (\pi - \omega)}}.
\label{qjdef}
\end{equation}
Note that the argument of the exponential in eq.(\ref{qjdef}) is 
different from that in eq.(\ref{omeq1}) because the original integral,
that is the lower line of eq.(\ref{omeq1}), was multiplied by a
factor of $2\epsilon a e^{-\epsilon a}$ in order to obtain
$\Psi_j$ in eq.(\ref{omeq2}).

To calculate an approximation to the integral in eq.(\ref{omeq2}), we
expand the argument of the exponential, noting that \( a\rho \leq 1 \).
The first term in the power-series expansion of the arcsine cancels
with \( a\rho \), and the next term is in \( (a\rho)^{3} \) which we ignore,
such that
\begin{equation}
e^{-\epsilon (a + \arcsin (a\rho) -a\rho)} \simeq
e^{-\epsilon a (1 + (a\rho)^{3}/(6a))} \simeq
e^{-\epsilon a },
\label{dirtyprox}
\end{equation}
which is independent of \( \rho \), and can be moved outside the 
integral, which has now been reduced to the integration of the
function \( g(\rho) \) from eq.(\ref{gdef}). The integration of
\( g(\rho) \) breaks down into six integrals, five of which can be solved in
terms of elementary functions and the 
sixth in terms of the Riemann \( \Phi (z,s,v) \)
function \cite{gr65}. A full form for the solution, \( \Psi \),
appears in Appendix~A.

\subsection{Rays which avoid the cavity}

A ray which does not enter the cavity obeys the same propagation
equation, eq.(\ref{ray1shell1}), but has a different upper limit
on the distance, \( s \). Instead of integrating to the cavity
boundary, we integrate to the mid-point, where the radial
vector in perpendicular to the direction of the ray. At this point,
the value of \( s = R_{2} \cos \omega \), and the corresponding
radius is \( R_{2} \sin \omega \). We can therefore still use
eq.(\ref{omeq1}), eq.(\ref{p1}) and eq.(\ref{p2}), but noting that
the expression in eq.(\ref{p2}) now reduces to \( P = R/(R_{2} \sin
\omega )\), rather than \( 1 \). The integrations and approximations
which follow still apply, providing we use the new value of
\( P \).

\subsection{Outward bound rays}

Provided that we are interested only in emergent rays, we can
use the symmetry of the nebula to simplify matters greatly here.
The integral along the ray, \( \Psi_{j} \), is indentical, whether
we are integrating inward to the midpoint (or cavity), or
outward again towards the edge of the nebula. These integrals
become formally the same because when we convert from \( s' \)
to \( r \) as the integration variable, we must use different
roots of the expression
\begin{equation}
s' = R_{2} \cos \omega \pm (r^{2} - R_{2}^{2} \sin^{2} \omega )^{1/2},
\label{sroot}
\end{equation}
with the negative root required for the inward segment, and the
positive root holding for the outward ray. The result is that
the integrals over radius are identical. The form of
eq.(\ref{ray1shell1})
which applies to outward rays is
\begin{equation}
\delta I(s) = \delta I(s_{0}) +
e^{-h(s)}\int_{s_{0}}^{s} \sigma^{c}(r(s')) J(r(s')) 
e^{h(s')} ds',
\label{ray1out}
\end{equation}
where \( s_{0} \) is now either the midpoint, for a ray which
avoids the cavity, or the outward cavity boundary. We assume
in eq.(\ref{ray1out}) that there is negligible scattering within
the cavity. The input, \( \delta I (s_{0}) \), is just the
inbound solution, evaluated at the midpoint or the inbound
cavity boundary. If we define,
\begin{equation}
\zeta = \sigma^{c}(R) R^{2} j_{0}/(4 \epsilon a),
\label{zetadef}
\end{equation}
then we can re-write eq.(\ref{ray1out}) as
\begin{eqnarray}
\delta I(s) & = & \zeta \Psi_{j} \left\{
 e^{\epsilon \arcsin [(R_{2}/r(s_{0})) \sin \omega]}\right. \nonumber \\
            &  + & \left.
 e^{\epsilon \arcsin [(R_{2}/r(s)) \sin \omega}
                          \right\},
\label{ray2out}
\end{eqnarray}
noting that any constants in the exponentials have been cancelled
between the \( e^{-h} \) term outside the integrals, and the 
\( e^{h} \) inside. The equivalence of the integrals, \( \Psi_{j} \),
has been used, from the symmetry argument above. We can now
write down two versions of eq.(\ref{ray2out}). The first is for
a ray which avoids the cavity, where \( r(s_{0}) = R_{2} \sin \omega \)
and \( r(s) = R_{2} \) on emerging from the cavity. In this case,
\begin{equation}
\delta I(s) = \zeta \Psi_{j} e^{\epsilon \pi /2}
( 1 + e^{-\epsilon (\pi / 2 - \omega )} ).
\label{out_nocav}
\end{equation}
For the other case, where the ray crosses the cavity, the first
value changes to \( r(s_{0}) = R \); the second remains the same,
so we have,
\begin{equation}
\delta I(s) = \zeta \Psi_{j}
( e^{\epsilon \arcsin a} + e^{\epsilon \omega} ),
\label{out_cav}
\end{equation}
where the integral \( \Psi_{j} \) is given in Appendix~A (with 
appropriate limits).
\begin{figure}
\vspace{0.3cm}
\psfig{file=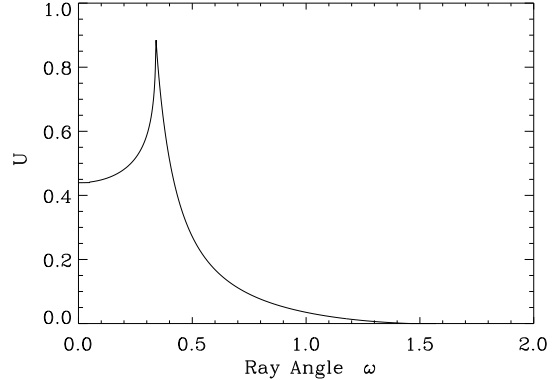,angle=0,width=8.0cm}
\caption{
The function \( U = \delta I(s)/(2 j_{0}R) \) for complete
path lengths \( s = s(R_{2}) \)
plotted as a function of entry angle into the nebula, \( \omega \).
The optical depth parameter is \( \chi^{c}(R) R = 1.38 \) as before, and
the ratio of the shell to cavity radii is \( R_{2}/R = 3.0 \).
The ratio of the scattering coefficient to the extinction
coefficient is \( \sigma^{c}(R)/\chi^{c}(R) = 0.25 \).
}
\label{scatplot}
\end{figure}

In Fig.~\ref{scatplot}, we plot the emergent specific intensity
for scattering, \(U=\delta I(s)/(2j_{0}R) \) over the full 
range of entry angles to the
nebula. The value of \(s\) is the largest path
length through the shell material for each angle.
The same optical depth parameter for extinction, \( 1.38 \), has
been used as in Fig.~\ref{specplot} and Fig.~\ref{jshellplot}.
The ratio of the outer (shell) radius to the inner (cavity)
radius is \( 3.0 \): this has not been used as a formal parameter
before, but the same value was also adopted for the above figures.
The ratio of the scattering coefficient to the extinction
coefficient has, rather arbitrarily, been set to \( 0.25 \).
For the perturbative method to be accurate, this value should
really be small, but is unlikely to be so for typical dust
models (see discussion in Section~\ref{pertstuff}). 
This parameter anyway
acts as a simple scaling factor which does not change the
shape of the function in Fig.~\ref{scatplot}. The normalising
factor of \( 2j_{0}R \) is the same as in the other specific
intensity plot, Fig.~\ref{specplot}.

The marked cusp in the curve in Fig.~\ref{scatplot} is at the
expected angle, \( \arcsin (R/R_{2}) \), where the ray
switches from paths which enter the cavity to paths which do not.
The path at this angle also corresponds to the maximum
distance travelled through the shell material. The solution
at smaller angles than that corresponding to the cusp comes
from eq.(\ref{out_cav}); at larger angles, eq.(\ref{out_nocav})
has been used.

\section{Solution for Emergent rays}

To use the symmetry property of the integral \( \Psi_{j} \), we
have already evaluated the perturbation, \( \delta I \) as
an emergent quantity above. In general, the complete solution,
for a given emergent ray, is given by,
\begin{equation}
I_{em}(s) = I(s) + \delta I(s),
\label{esoln}
\end{equation}
where \( s = 2 R_{2} \cos \omega \). For a ray which crosses
the cavity, \( I(s) \) comes from a form of eq.(\ref{combsoln})
where \( r = R_{2} \) and \( \theta = \omega \), whilst
\( \delta I(s) \) is given by eq.(\ref{out_cav}). Overall,
\begin{equation}
I_{em}\! = \! 2j_{0}R(1-a^{2})^{1/2} \!
 e^{-\epsilon (\arcsin a - \omega )}
       + \zeta \Psi_{j}
( e^{\epsilon \arcsin a} + e^{\epsilon \omega} ).
\label{esoln_cav}
\end{equation}
A ray which does not cross the cavity has only the \( \delta I \)
contribution, and is given by eq.(\ref{out_nocav}) without
modification.

\subsection{The observer's view of the model}

To an observer with perfect angular resolution, an emergent ray of
given exit angle, \( \omega \), characterises a circular strip of
a spherical surface. This strip has area \( 2\pi R_{2}^{2} \sin u \),
where \( u \) is the polar angle measured from the line of sight
towards the limb of the nebula. The observer cannot see this surface
as a whole, but only a 2-D projection of it. The projected area
of the strip is
\begin{equation}
dA_{\perp} = 2\pi R_{2}^{2} \sin u \cos u du
\label{striparea}
\end{equation}
It is straightforward to see that for any given strip, the
polar angle \( u \) is identical to the entry/exit angle of the
ray, \( \omega \), which has allowed values between
\( 0 \) and \( \pi / 2 \). The observer with perfect
angular resolution will therefore be able to pick out a small
piece of a given (projected) strip, and will measure the
associated specific intensity, \( I_{em} \), as a brightness.
For a nebula of radius \( R_{2} \), at a distance \( d \) from
the observer, the measured brightness at an angle \( \Theta \)
from the centre of the nebula, as seen by the observer, corresponds
to exit angle \( \omega = \arcsin ( \Theta d / R_{2} ) \). The
brightness itself can be found by subsitituting this angle
into either eq.(\ref{esoln_cav}) or eq.(\ref{out_nocav}).

The other extreme observer's view is that of a telescope with
a beam that is much larger than the nebula. In this case, the
specific intensity cannot be measured directly, and a flux,
averaged over the whole object, is obtained instead. To a very
good approximation, this flux is given by
\begin{equation}
F = \frac{2\pi R_{2}^{2}}{d^{2}}
    \int_{0}^{\pi /2} I(\omega ) \sin \omega \cos \omega d\omega,
\label{obsflux}
\end{equation}
where, as previously, the functional form of \( I(\omega ) \), is
taken from eq.(\ref{esoln_cav}) or eq.(\ref{out_nocav}),
depending on whether or not the ray at angle \( \omega \)
traverses the cavity.

\section{Fits to NGC6537}
\label{txt_parms}

The function summarised in eq.(\ref{esoln}) and eq.(\ref{esoln_cav})
was fitted to observational data describing the brightness variation
of the H$\alpha$ and H$\beta$ spectral lines as functions of angular
position across the nebula. The fits were made with respect to
four variable parameters in the theoretically-derived function:
$x=R/R_2$, the ratio of the cavity radius to the overall radius of
the nebula; $S=\sigma^c/\chi^c$, the ratio of the scattering to
extinction coefficients in the continuum;
$\tau = R \chi^c(R)$, the optical depth parameter,
%corresponding to traversing
%a distance equal to the cavity radius
%through material with an extinction
%coefficient calculated at the cavity radius, 
and an overall intensity-axis scale factor,
$Y$, allowing a fit to normalized observational data. The optimum
fit for each observational data set was taken to be that 
with the minimum value of the $\chi^2$-statistic,
\begin{equation}
\chi^2 = \frac{1}{N-4} \sum_{i=1}^N \left(
\frac{I_i - I_{em}(\sin \omega_i,x,S,\tau,Y)}
     {\sigma_i}
                                   \right)^2,
\label{eq_chisq}
\end{equation}
for a data set with $N$ entries of the form $(\sin \omega_i,I_i)$, and
the four free parameters introduced above.
Details of the
treatment of the observational data, and of the fitting function,
are described below.

\subsection{Data from NGC6537}
\label{datangc6537}

Data were extracted from Hubble Space Telescope H$\alpha$ and
H$\beta$ images \cite{Matsuura05}.
Seven straight-line slices were taken through the
nebula, each at a different angle on the sky. For each slice,
data were recorded for both H$\alpha$ and H$\beta$, and for each
slice and spectral line,
data were organised into pairs consisting
of a coordinate position along the slice,
and a corresponding specific intensity. The following operations
were applied to convert these data into a form suitable for
fitting: For each slice and line, two pixel positions were
found for the emission peaks, corresponding to the intersections
of the slice with the cavity boundary. The coordinate origin
was then shifted to the mid-point of the
peak positions, and the absolute value of the slice coordinate
taken, transforming the data to a pair of radial brightness profiles
measured from an origin at the approximate centre of the
nebula. An approximate correction for the aspherical nature
of the nebula was imposed by re-scaling the radial axis
such that the origin to peak distance was the same for 
all radial profiles. The outer radius of
the nebula was taken to be the smallest maximum value on the
new scale, and the remaining profiles truncated to the
same distance. Finally, the new radial coordinate was re-scaled
once more to the range $0.0-1.0$, with $1.0$ corresponding to the
now common outer radius. We note that the H$\beta$ data covered a
larger angular range on average than the H$\alpha$ data, leading
to a larger value of $R_2$ in H$\beta$ by a factor of $1.51$, and
a different radial scaling for the two lines. 

For individual radial profiles, the specific intensities were
scaled such that the peak brightness was set to $1.0$. Mean
profiles for each spectral line were also constructed by averaging
over all the un-normalized
individual brightness profiles (on the normalised radial
scale for each line
defined above), and then normalising the averaged brightnesses
to a peak height of $1.0$. 
 
In the averaged profiles, the standard errors resulting from the
averaging process far exceeded the formal measurement errors
in the observational data. The former were calculated at the
peak position to be $0.098$ for H$\alpha$ and $0.054$ for H$\beta$,
as fractions of the mean peak height. The latter were only of
order $1.7\times 10^{-4}$ and $3.0\times 10^{-4}$, respectively.
For the normalised data sets, with peak heights of $1.0$ we assumed
absolute uncertainties at all radial points to be equal to those
at the respective peaks: $0.098$ for H$\alpha$ and $0.054$ for
H$\beta$. The same absolute errors were assumed for the fits
to the individual radial profiles. Values of
$N-4$ were $75$ for H$\alpha$ and $120$ for H$\beta$.

\subsection{Parameter ranges}

The parameter $x$ has the formal range $0.0-1.0$, but a much
smaller realistic range can be selected for any radial profile
by observing the position of the scattering peak (see also
Fig.~\ref{scatplot}). Both the optical depth parameter, $\tau$, and the
scale-factor, $Y$, have the possible range $0.0-\infty$. The
most problematic parameter is $S$, the ratio of the continuum
scattering to extinction coefficients. The scattering has been
treated as a perturbation in Section~\ref{sectpert}, 
so strictly speaking
only the range $0.0-0.1$ is available to $S$. However, from dust
models, we expect the ratio of the scattering to absorption 
cross-sections to be of order $2-3$ (corresponding to
$S=0.666-0.75$) in the optical region, as already discussed in
Section~\ref{pertstuff}. We have therefore fitted each profile
twice: once for a true perturbation, with $S$ limited to the
range $0.0-0.1$, and once with the range $0.0-0.666$, with
the upper limit dictated by computed dust parameters. For
each spectral line, we have fitted the averaged profile, and
one selected individual profile. The fitting parameters for
H$\alpha$ and H$\beta$ have been taken to be entirely independent
in this preliminary study.

\subsection{Results of fitting}

\begin{figure*}
\vspace{0.0cm}
\psfig{file=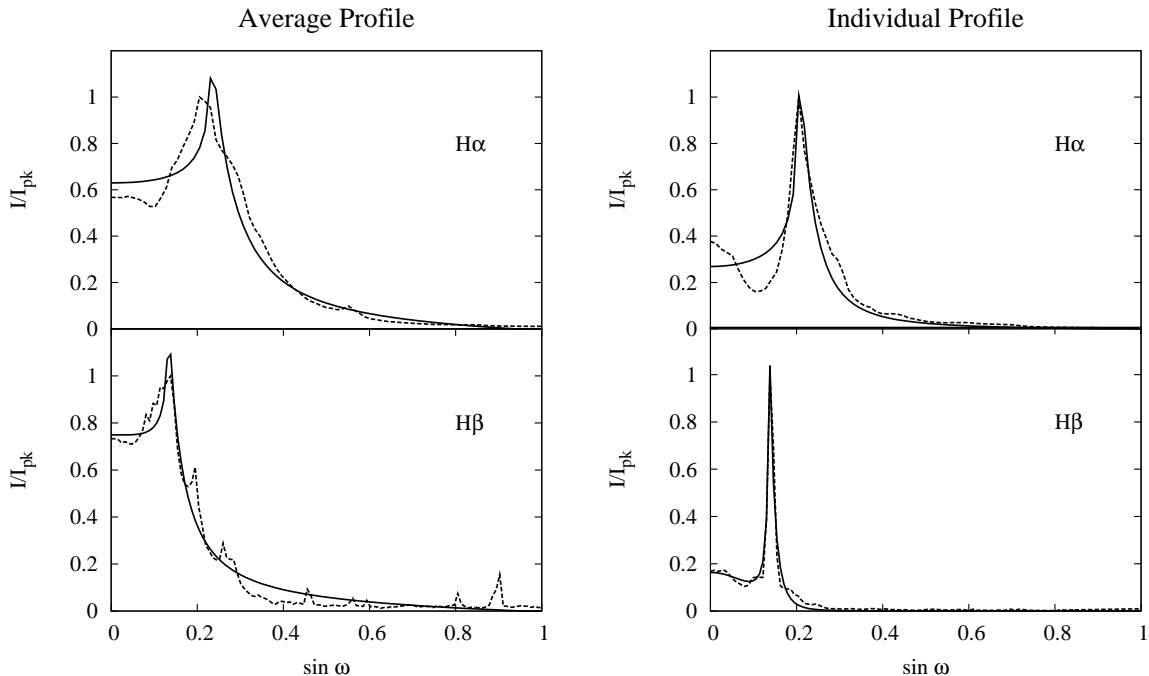,angle=0,width=18.0cm}
\caption{
Best fit profiles (solid lines) following the functional forms
in eq.(\ref{esoln}) and eq.(\ref{esoln_cav}) to observational
data from NGC6537 (dashed lines). The left-hand panel uses
the observational data averaged over $14$ radial profiles, whilst
the right-hand panel uses just one individual profile. For each
panel, the upper graph is for the H$\alpha$ line, and the lower,
for H$\beta$.
}
\label{fullfits}
\end{figure*}

In Fig.~\ref{fullfits} we show the results of the fits where the allowed
range of $S$ is $0.0-0.666$. The upper graphs are fits to
H$\alpha$ data, and the lower graphs, to H$\beta$. For each
line, the left-hand panels are fits to the averaged profile,
whilst the right-hand panels show fits to a selected individual
profile. Note that the individual H$\beta$-fit (bottom right) is
the only case in which the best fit fell in the perturbation
range. A general difficulty with fits to averaged data is the
broadening of the scattering peak in the averaging process, a
consequence of the asphericity of the nebula that has only been
approximately mitigated by the rescaling operations described
in Section~\ref{datangc6537}. A full table of the fitting
parameters and quality estimates appears in Table~\ref{tabfit}.
%Put the first fit figure here.
%Fit parameters table
\begin{table}
\caption{Values of the fitting parameters, $x$,$S$,$\tau$ and
$Y$ (see Section~\ref{txt_parms}), that gave the best value of
$\chi^2$, as defined by eq.(\ref{eq_chisq}), for the associated data set. 
Type `Avg' refers to averaged data, type `Ind', to an individual
profile. The symbol $S'$ has the value $3.2\times 10^{-4}$. Note
that values of $x$ for H$\beta$ should be multiplied by $1.51$ to
place them on the same radial scale as those of H$\alpha$.
}
\label{tabfit}
\begin{tabular}{@{}lrrrrrr}
\hline
Line/Type     & $S_{max}$& $x$     & $S$  & $\tau$  & $Y$    &$\chi^2$ \\
\hline
H$\alpha$/Avg & $0.666$ & $0.2326$&$0.666$&$1.665$ &$0.372$ &$0.491$  \\
H$\alpha$/Avg & $0.100$ & $0.2120$&$0.100$&$2.413$ &$0.099$ &$1.423$  \\
H$\alpha$/Ind & $0.666$ & $0.2060$&$0.585$&$2.850$ &$0.160$ &$0.384$  \\
H$\alpha$/Ind & $0.100$ & $0.2060$&$0.100$&$3.175$ &$0.785$ &$0.434$  \\
H$\beta$/Avg  & $0.666$ & $0.1320$&$0.300$&$1.602$ &$0.750$ &$0.755$  \\
H$\beta$/Avg  & $0.100$ & $0.1325$&$0.100$&$1.995$ &$1.818$ &$1.079$  \\
H$\beta$/Ind  & $0.666$ & $0.1346$&$S'   $&$6.193$ &$26.10$ &$0.146$  \\
\hline
\end{tabular}
\end{table}

\begin{figure*}
\vspace{0.0cm}
\psfig{file=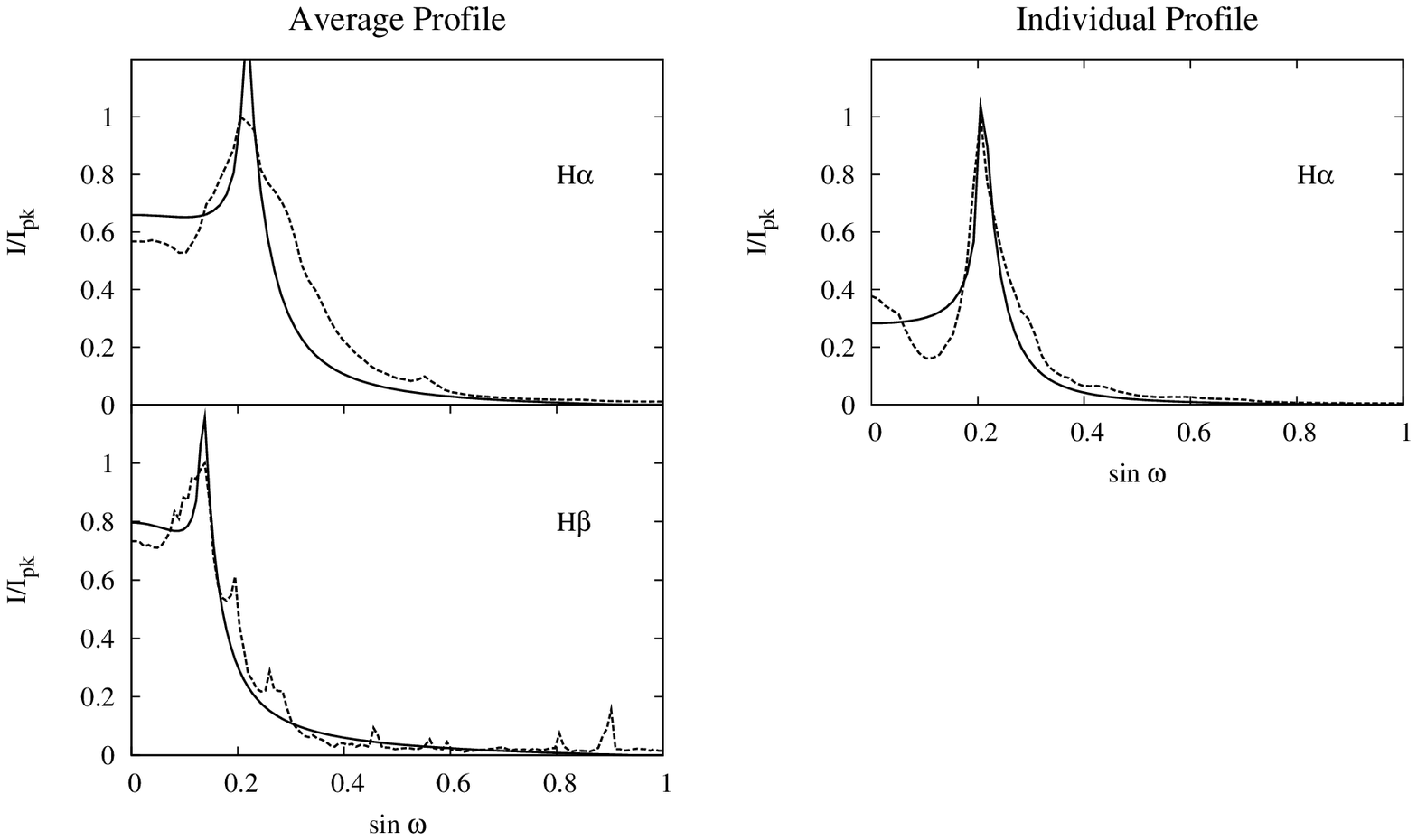,angle=0,width=18.0cm}
\caption{
As for Fig.~\ref{fullfits} except that perturbation values
of $S$ are enforced ($S_{max} = 0.1$). The individual H$\beta$
profile (omitted here) is identical to that in Fig.~\ref{fullfits}.
}
\label{pertfits}
\end{figure*}

Figure~\ref{pertfits} shows the fits in which $S$ was restricted to the
perturbation range, $0.0-0.1$. Only in the case of the fit
to the selected H$\beta$ profile was the best overall fit
found in the perturbation range, and this fit has already
been shown in Figure~\ref{fullfits}. We note that this particular
fit has rather peculiar parameters (see Table~\ref{tabfit}), with
a much larger optical depth parameter (and much smaller scattering
parameter, $S$) than the others.
%Put the perturbation fits figure here.

The easiest parameter to compare with observations is $x$.
Values of $x=R/R_2$ are markedly different for the two spectral lines,
but this is simply a consequence of the greater angular extent
of the data (larger $R_2$) for H$\beta$. When corrected to the
value of $R_2$ for H$\alpha$, by multiplying by $1.51$, the values of
$x$ for H$\beta$, in the order they appear in Table~\ref{tabfit},
are $0.1993, 0.2001$ and $0.2032$. These values are then consistent
with those found for H$\alpha$.
%with a significantly smaller apparent cavity for H$\beta$. The
%different upper limits on $S_{max}$ produce little
%difference in the optimum values of $x$, and 
%$x=R/R_2=0.22$ is a useful average for H$\alpha$, but
%only $0.13$ for H$\beta$
For the purposes of the optical-depth comparison (see below), it
is useful to calculate uncorrected values of $1-x$ for both
lines, yielding 0.79 for H$\alpha$ and 0.87 for H$\beta$.

We now consider the
results for the optical depth parameter. 
In Section~\ref{txt_pmodneb}, the observed magnitude drops
across the nebula, owing to extinction, ranged from 1.5 to 2.2
magnitudes, corresponding to intensity ratios of 0.251 and 0.132
along a line of sight. The respective optical depths are
1.38 and 2.03. We note that the fitted values of $\tau$ in
Table~\ref{tabfit} are of the optical depth parameter introduced
at the end of Section~\ref{combinedinshell}, and that this is
related to the optical depth of a radial ray passing through the
full thickness of the nebula by $\tau_r = \tau (1-x)$ (see
eq.(\ref{eq_thelastone}) with the limiting value of
$r \rightarrow R_2$). Fitted
values of $\tau$ from the averaged data
of 1.67 for H$\alpha$ and 1.60 for H$\beta$ 
with the full range of $S$ then yield respective optical depths of
1.32 and 1.39. If  the perturbation restriction
is enforced, the optical depths are 1.90 and 1.74, results that are
reasonably consistent with the observational values
Note that considerably different optical depths are
recovered from the selected individual profile, suggesting a
locally clumpy medium.

\section{Conclusion}

An approximate analytical function has been derived that
predicts the brightness of a spectral line, within some
filter bandwidth, as a function of angular distance from the
centre of a cavity/shell planetary nebula. The scattering part
of this function is derived as a perturbation on the radiative
transfer solution. This function has
been fitted to observational data from NGC6537 in the H$\alpha$
and H$\beta$ lines. The best-fit optical depth parameter in the model
gives reasonable agreement with the observed optical extinction
in the nebula. However, best-fit values for the scattering
parameter are mostly too high to be consistent with a true
perturbation. Consistent values were also found for the sizes
of the cavity in both lines when the outer radius for H$\alpha$ was
applied to both spectral lines. The mean of the values of
$x$ from Table~\ref{tabfit} was $0.21$, and the mean of
the adjusted values for H$\beta$ was $0.20$.

The analytical function may also have applications to other sources with a
hot, largely ionized, interior, surrounded by a shell rich in
dust. Examples include post-AGB stars, supernova remnants,
symbiotic stars, quasars, and highly-magnetized planets in
addition to the planetary nebulae discussed here.

\subsection*{ACKNOWLEDGMENTS}

MDG acknowledges STFC (formerly PPARC) for financial support under
 the 2005-2010 
rolling grant, number PP/C000250/1.

\appendix

\section{The Integral of \( g(p) \)}

The solution shown here is for any ray.
For a ray which enters the cavity, the upper limit of these
integrals should be set to \( P=1 \).
For a ray which avoids the cavity, the upper limit
should be \( R/(R_{2}\sin \omega )\). The
integral of \( g(p) \) in eq.(\ref{gdef}) breaks down into
six integrals. The first two, trivially, are
\begin{equation}
I_{1} = 2\epsilon a \int_{R/R_{2}}^{P} d\rho = 2\epsilon a ( P - R/R_{2})
\label{gi1}
\end{equation}
and
\begin{equation}
I_{2} = \int_{R/R_{2}}^{P} d\rho/\rho = 2\ln (R_{2}P/R)
\label{gi2}
\end{equation}
The third can be integrated by parts, with the logarithm as the
differentiated part. The result is,
\begin{eqnarray}
I_{3} & = & \int_{R/R_{2}}^{P} \ln \left(
         \frac{1+\rho}{1-\rho}
                               \right) d\rho \nonumber \\  & = &
(1+P)\ln (1+P) + (1-P)\ln (1-P) \nonumber \\
&  - & \frac{R}{R_{2}} \ln \left(
           \frac{R_{2} + R}{R_{2} - R}
                             \right) -
 \ln \left(
 1 - \frac{R^{2}}{R_{2}^{2}}
     \right).
\label{gi3}
\end{eqnarray}

The fourth integral is the most problematic, as the result
cannot be expressed in terms of elementary functions. If the
integral is split, so that
\begin{eqnarray}
\frac{I_{4}}{\epsilon a} & = & \int_{R/R_{2}}^{P} \frac{1}{\rho} \ln \left(
         \frac{1+\rho}{1-\rho}
                               \right) d\rho \nonumber \\
      & = &
  \int_{R/R_{2}}^{P} \frac{\ln (1+\rho) }{\rho} d\rho -
  \int_{R/R_{2}}^{P} \frac{\ln (1-\rho) }{\rho} d\rho,
\label{gi4a}
\end{eqnarray}
then both the expressions on the bottom line of eq.(\ref{gi4a})
conform to a standard integral \cite{gr65}, which is their 
no.~2.728 (version~2). The solution is written in terms of the
function \( \Phi (z,s,v) \), which is discussed in detail in
Section~9.55 of Gradshteyn \& Ryzhik \shortcite{gr65}. The
final result is
\begin{eqnarray}
I_{4} & = & \epsilon a \left\{ P [
           \Phi (-P,2,1) + \Phi (P,2,1)]
                   \right. \nonumber \\
      & - & \left.
             \frac{R}{R_{2}} \left[
           \Phi(-\frac{R}{R_{2}},2,1) +
           \Phi(\frac{R}{R_{2}},2,1)
                           \right]
                   \right\}.
\label{gi4b}
\end{eqnarray}

The fifth integral can be solved in a similar way to the simpler
\( I_{3} \), integrating by parts with the logarithm as the
differentiated part. The integrated part of the result can be
solved by an expansion in partial fractions, and the overall
expression is
\begin{eqnarray}
I_{5}& = & \int_{R/R_{2}}^{P} \frac{1}{\rho^{2}} \ln \left(
         \frac{1+\rho}{1-\rho}
                               \right) d\rho \nonumber \\ & = &
\frac{R_{2}}{R} \ln \left(
      \frac{R_{2}+R}{R_{2}-R}
                    \right) + 
                \ln \left(
     \frac{R_{2}^{2}-R^{2}}{R^{2}}
                    \right) \nonumber \\ & + &
2\ln P -(\frac{1}{P}+1)\ln (1+P) + (\frac{1}{P}-1) \ln(1-P).
\label{gi5}
\end{eqnarray}
The final integral can be solved in a broadly similar manner to
\( I_{5} \), and the solution is
\begin{eqnarray}
\frac{I_{6}}{\epsilon a}& = & \int_{R/R_{2}}^{P} \rho
\ln \left(
         \frac{1+\rho}{1-\rho}
                               \right) d\rho \nonumber \\ & = &
\frac{1}{2} \left(
1-\frac{R^{2}}{R_{2}^{2}}
\right)
  \ln \left(
       \frac{R_{2}+R}{R_{2}-R}
      \right) + P - \frac{R}{R_{2}} \nonumber \\ 
& + &
\frac{1}{2} (1-P^{2})
  \ln \left(
    \frac{1-P}{1+P}
      \right).
\label{gi6}
\end{eqnarray}

The integral \( \Psi_{j} \), in its
approximate analytical form, is now given by eq.(\ref{omeq2}), with
its lower line replaced by the expression,
\begin{equation}
e^{-\epsilon a} \sum_{k=1}^{6} I_{k},
\label{guff}
\end{equation}
where the \( I_{k} \) are the integrals in eq.(\ref{gi1}) to
eq.(\ref{gi6}). The necessary approximation to the exponential
is discussed in Section~8 (see eq.(\ref{dirtyprox})).


\begin{thebibliography}{}
\bibitem[\protect\citename{Abramowitz \& Stegun }1965]{absteg65}
Abramowitz M., Stegun I.M., 1965, Handbook of Mathematical
Functions, Dover Publishing, New York, 8th Dover printing
\bibitem[\protect\citename{Balick \& Frank }2002]{Balick02}
Balick B., Frank A., 2002, ARA\&A, 40, 439
\bibitem[\protect\citename{Bilikova et al. }2007]{Bilikova07}
Bilikova J., Williams R.N.M., Chu Y.-H., Gruendl R.A., Lundgren B.F.,
2007, AJ, 134, 2308
\bibitem[\protect\citename{Dwek et al. }1987]{Dwek87}
Dwek E., Hauser M.G., Dinerstein H.L., Gillett F.C., Rice W.L.,
1987, ApJ, 315, 571
\bibitem[\protect\citename{Ercolano, Barlow \& Storey }2005]{Ercolano05}
Ercolano B., Barlow M.J., Storey P.J., 2005, MNRAS, 362, 1038
\bibitem[\protect\citename{Gradshteyn \& Ryzhik }1965]{gr65}
Gradshteyn I. S., Ryzhik I.M., 1965, Table of Integrals Series
and Products, Academic Press, 4th edition.
\bibitem[\protect\citename{Gray \& Field }1995]{gra95}
Gray M. D., Field D., 1995, A\&A, 298, 243
\bibitem[\protect\citename{Kwok, Purton \& Fitzgerald }1978]{Kwok78}
Kwok S., Purton C.R., Fitzgerald P.M., 1978, ApJ, 219, 125
\bibitem[\protect\citename{Lagage et al. }1996]{Lagage96}
Lagage P.O., Claret A., Ballet J., Boulanger F., Cesarsky C.J.,
Cesarsky D., Fransson C., Pollock A., 1996, A\&A, 315, 273
\bibitem[\protect\citename{Matsuura et al. }2005]{Matsuura05}
Matsuura M., Zijlstra A.A., Gray M.D., Molster F.J., Waters L.B.F.M.,
2005, MNRAS, 363, 628
\bibitem[\protect\citename{Ossenkopf, Henning \& Mathis }
1992]{ossenkopf92}
Ossenkopf V., Henning Th., Mathis J.S., 1992, A\&A, 261, 567
\bibitem[\protect\citename{Peraiah }2002]{per02}
Peraiah A., 2002, An Introduction to Radiative Transfer, CUP,
Cambridge
\bibitem[\protect\citename{Perinotto et al. }2004]{perinotto04}
Perinotto M., Sch\"{o}nberner D., Steffen M., Calonaci C., 2004,
A\&A, 414, 993


\end{thebibliography}
\end{document}